\documentclass[draftcls, onecolumn, 12pt, twoside]{IEEEtran}
\usepackage[cmex10]{amsmath}%[cmex10]
\usepackage{amssymb}
\usepackage{graphicx}
\usepackage{subfigure}
\usepackage{cite}
\usepackage{subeqnarray}
\usepackage{diagbox}
\usepackage{booktabs}
\usepackage{multirow}
\usepackage{algorithm}
\usepackage{algorithmic}
\usepackage{url}
\usepackage{xcolor}
\usepackage{array}
\usepackage{footnote}
\makesavenoteenv{tabular}

\newtheorem{theorem}{Theorem}

\newtheorem{proposition}[theorem]{Proposition}
\newtheorem{corollary}{Corollary}
\newtheorem{definition}{Definition}
\newtheorem{example}{Example}
\newtheorem{remark}{Remark}
\newtheorem{strategy}{Strategy}

\begin{document}

\title{The New Multi-Edge Metric-Constrained PEG/QC-PEG Algorithms for Designing the Binary LDPC Codes With Better Cycle-Structures}

\author
{
Xuan~He,~\IEEEmembership{Student Member,~IEEE,}
Liang~Zhou,~%\IEEEmembership{Member,~IEEE,}
and~Junyi~Du,~\IEEEmembership{Student Member,~IEEE}% <-this % stops a space
\thanks{Part of this work \cite{	He15} has been published in the 2015 IEEE International Symposium on Information Theory.}% <-this % stops a space
%\thanks{Manuscript received April 19, 2005; revised September 17, 2014.}
}

\markboth{He \MakeLowercase{\emph{et al.}}: The MM-PEGA/MM-QC-PEGA design the LDPC codes with better cycle-structures}
{He \MakeLowercase{\emph{et al.}}: The MM-PEGA/MM-QC-PEGA design the LDPC codes with better cycle-structures}

\maketitle

\begin{abstract}

To obtain a better cycle-structure is still a challenge for the low-density parity-check (LDPC) code design.
This paper formulates two metrics  firstly so that the progressive edge-growth (PEG) algorithm and the approximate cycle extrinsic message degree (ACE) constrained PEG algorithm are unified into one integrated algorithm, called the metric-constrained PEG algorithm (M-PEGA).
Then, as an improvement for the M-PEGA, the multi-edge metric-constrained PEG algorithm (MM-PEGA)  is proposed based on two new concepts, the multi-edge local girth and the edge-trials.
The MM-PEGA with the edge-trials, say a positive integer $r$, is called the $r$-edge M-PEGA, which constructs each edge of the non-quasi-cyclic (non-QC) LDPC code graph through selecting a check node whose $r$-edge local girth is optimal.
%, instead of through selecting a check node whose one-edge local girth is optimal as that in the M-PEGA.
In addition, to design the QC-LDPC codes with any predefined valid design parameters, as well as to detect and even to avoid generating the undetectable cycles in the QC-LDPC codes designed by the QC-PEG algorithm, the multi-edge metric constrained QC-PEG algorithm (MM-QC-PEGA) is proposed lastly.
It is verified by the simulation results that increasing the edge-trials of the MM-PEGA/MM-QC-PEGA is expected to have a positive effect on the cycle-structures and the error performances of the LDPC codes designed by the MM-PEGA/MM-QC-PEGA.

\end{abstract}


\IEEEpeerreviewmaketitle

%%%%%%%%%%%%%%%%%%%%%%%%%%%%%%%%%%%%%%%%%%%%%%%%%%%%%%%%%%%%%%%%%%%%%%%%%%%%%%%%%%%%%%%%%%%%%%%%%%%%%%%%%%%%%%%%%%%%%%%%%%%%%%%%%%%%%

\section{Introduction}

%\IEEEPARstart{L}{ow-density} parity-check (LDPC) code \cite{	Gallager62} has been an attractive research topic in coding theory over the past decade due to its near capacity-approaching performances and successful practical applications.
%To achieve a low decoding threshold and a good error performance in the waterfall region, Richardson \emph{et al.} \cite{	Richardson01a, Richardson01b} developed the density evolution (DE) to optimize the degree distribution of the LDPC code.
%Meanwhile, the very long DE-optimized LDPC codes approach the Shannon limit very closely \cite{	Chung01}.
%However, DE should not be the only design criterion, as short and medium length LDPC codes contain short inevitable cycles, which significantly deteriorate the error performances of the LDPC codes \cite{	Hu05, Tian04}.

\IEEEPARstart{T}{he} error floor performance of the low-density parity-check (LDPC) code \cite{	Gallager62} over the additive white Gaussian noise (AWGN) channel is closely related to a cycle-structure called the trapping set (TS) \cite{	Richardson03}.
However, it was shown in \cite{	McGregor10} that to find or even to approximate the minimum size of the TS in a Tanner graph (TG) \cite{    Tanner81} is NP-hard.
%, not to mention the hardness of that to improve the error performance of the LDPC code through directly optimizing its TSs.
%Recently, some efficient algorithms for finding the elementary trapping sets (ETSs) \cite{	Richardson03} were proposed in \cite{Karimi12, Karimi14,	Hashemi15}.
%In addition, the LDPC code design algorithms in \cite{	Zheng10, Asvadi11, Khazraie12, Diouf15} try to optimize the ETSs real timely during the construction processes of the LDPC codes.
In addition, the methods in \cite{	Zheng10, Asvadi11, Khazraie12, Diouf15}, which try to directly optimize the elementary trapping sets (ETSs) \cite{	Richardson03} real timely during the construction process of the LDPC codes, hold a high realization complexity as well as a high computational complexity, and sometimes will fail the construction.
With regard to this, some other widely used cycle-structures, such as the girth \cite{	Hu05}, the approximate cycle extrinsic message degree (ACE) \cite{	Tian04, Xiao04}, and the ACE spectrum \cite{	Vukobratovic07,	Vukobratovic08,	Vukobratovic09, Asvadi12}, which can be calculated simply and efficiently, have thus been used in this paper for the design and the analysis of the LDPC code.

Since good cycle-structures and good error performances have been achieved by the progressive edge-growth (PEG) algorithm \cite{	Hu05} as well as the ACE constrained PEG algorithm \cite{	Vukobratovic08}, while both of the algorithms hold a low realization complexity and a polynomial computational complexity, and will never fail the construction, thus our work is closely related to the ideas in \cite{Hu05} and \cite{Vukobratovic08}.
To be specific, this paper formulates two metrics firstly so that the PEG algorithm \cite{	Hu05} and the ACE constrained PEG algorithm \cite{	Vukobratovic08} are unified into one integrated algorithm, called the metric-constrained PEG algorithm (M-PEGA).
Then, as an improvement for the M-PEGA, the \emph{multi-edge metric-constrained PEG algorithm (MM-PEGA})  is proposed based on two new concepts, the \emph{multi-edge local girth} and the \emph{edge-trials}.
The MM-PEGA with the edge-trials, say a positive integer $r$, is called the $r$-edge M-PEGA, which is implemented under the framework of the M-PEGA but adopts a different \emph{selection strategy}.
More precisely, the $r$-edge M-PEGA constructs each edge of the non-quasi-cyclic (non-QC) LDPC code graph through selecting a check node (CN) whose $r$-edge local girth is optimal, instead of through selecting a CN whose one-edge local girth is optimal as that in the M-PEGA.
It's illustrated that the one-edge M-PEGA is equivalent to the M-PEGA.
In addition, to calculate the multi-edge local girth, a depth-first search (DFS) \cite{	introAlgo01} like algorithm is proposed.
%Furthermore, a new method is developed to accelerate the MM-PEGA.

QC-LDPC codes are more hardware-friendly compared to other types of LDPC codes in encoding and decoding. Encoding of the QC-LDPC codes can be efficiently implemented using simple shift registers \cite{	Li06}. In addition, the revolving iterative decoding algorithm \cite{	Liu13,	Li14} significantly reduces the hardware implementation complexity of a QC-LDPC decoder. So, amount of researchers show great interest in the construction of the QC-LDPC codes. At the same time, well designed QC-LDPC codes perform as well as other types of LDPC codes \cite{	Kou01, Lan07, Li14, Myung05, Jiang09, Jiang14, Huang10, Asvadi11, Asvadi12, Park13, Mitchell14, O'Sullivan06, Bocharova12, Karimi13, Fossorier04, Li04, Lin08, Diouf15}.	

In \cite{ Kou01, Lan07, Li14}, construction methods for the QC-LDPC code based on the finite field were proposed. These methods usually firstly construct the complete check matrix of the QC-LDPC code, in which each circulant is a \emph{circulant permutation matrix (CPM)}, and then adopt the masking technique \cite{	Xu07} to adjust the degree distributions of both the variable nodes (VNs) and the CNs.
In \cite{ Myung05, Jiang09,	Jiang14}, the Chinese Remainder Theory (CRT) is adopted to accelerate the construction process of the QC-LDPC code, keeping the girth of the target LDPC code not smaller than that of the base matrix.
Methods in \cite{ Huang10, Asvadi11, Asvadi12, Park13, Mitchell14, O'Sullivan06, Bocharova12, Karimi13} construct the QC-LDPC code by carefully cyclically lifting the protograph. In order to achieve the desirable large girth, the constraints which have been derived in \cite{	Fossorier04} to ensure the corresponding girth must be satisfied. However, it's usually impossible to satisfy the constraints unless the protograph is very sparse and the lifting size is quite large.
%Meanwhile, some lower bounds of the lifting size have been reported in \cite{O'Sullivan06, Bocharova12, Karimi13, Fossorier04}, and a method for greedily minimizing the lifting size was proposed in \cite{	Huang10}.
Moreover, in \cite{	Kou01, Lan07, Li14, Myung05, Jiang09,	Jiang14, Huang10, Asvadi11, Asvadi12, Park13, Mitchell14, O'Sullivan06, Bocharova12, Karimi13, Fossorier04}, the design parameters, such as the size of the check matrix, the degree distribution of the check matrix, and the size of the circulant, are not as flexible as that in the QC-PEG algorithm (QC-PEGA) \cite{	Li04}.
%Besides, most QC-LDPC code design methods in \cite{	Kou01, Lan07, Li14, Myung05, Jiang09,	Jiang14, Huang10, Asvadi11, Asvadi12, Park13, Mitchell14, O'Sullivan06, Bocharova12, Karimi13, Fossorier04} require each nonzero circulant of the check matrix to be a CPM.
Instead, the QC-PEGA \cite{	Li04} is suitable for designing the QC-LDPC code with any predefined valid design parameters.
However, the QC-PEGA \cite{	Li04} sometimes results in 4-cycles (cycles of length 4) just in a single circulant of the check matrix when there contain multiple edges.
To avoid generating 4-cycles in a single circulant of the check matrix, the circulant-permutation-PEG algorithm (CP-PEGA) \cite{	Lin08} restricts each nonzero circulant of the check matrix to be a CPM during the construction process of the QC-PEGA \cite{	Li04}.
%But instead, when the weight of some column of the check matrix exceeds the number of circulants in the assumed column, the CP-PEGA \cite{	Lin08} will fail the construction.

To flexibly design the QC-LDPC codes with better cycle-structures, an improvement for the QC-PEGA \cite{	Lin08}, called the \emph{multi-edge metric-constrained QC-PEG algorithm (MM-QC-PEGA)}, is proposed in this paper.
On the one hand, the MM-QC-PEGA is implemented under the framework of the QC-PEGA \cite{Li04} so that it can construct the QC-LDPC codes with any predefined valid design parameters.
On the other hand, following the idea of the $r$-edge M-PEGA, the $r$-edge M-QC-PEGA constructs each edge of the QC-LDPC code graph through selecting a CN whose $r$-edge local girth is optimal.
As a result, the undetectable cycles in the QC-LDPC codes designed by the QC-PEGA \cite{Li04} and the CP-PEGA \cite{Lin08} become detectable and even avoidable in the codes designed by the MM-QC-PEGA.

To investigate the cycle-structure as well as the error performance, plenty of simulations have been performed over the binary LDPC codes, which are designed by the MM-PEGA, the MM-QC-PEGA, and some of the conventional LDPC code design algorithms respectively.
According to the simulation results, the proposed algorithms, i.e., the MM-PEGA and the MM-QC-PEGA, are more effective than the conventional design algorithms in terms of designing the LDPC codes with better cycle-structures and better error performances.
In addition, compared to the non-QC-LDPC codes designed by the MM-PEGA, the QC-LDPC codes, which are designed by the MM-QC-PEGA with the similar design parameters, sometimes achieve better cycle-structures and better error performances.

The rest of this paper is organized as follows.
Section \ref{sect: backgrounds} introduces some preliminaries, notations, and backgrounds.
Section \ref{sect: MM-PEGA} firstly defines the multi-edge local girths and the edge-trials, following which the MM-PEGA is proposed. Then, a DFS like algorithm is proposed to calculate the multi-edge local girths. %At the end of this section, a new method is developed to accelerate the MM-PEGA.
Section \ref{sect: MM-QC-PEGA} proposes the MM-QC-PEGA to design the QC-LDPC codes. Section \ref{sect: simulation results} presents some simulation results. And finally, Section \ref{sect: conclusions} concludes the paper.

%%%%%%%%%%%%%%%%%%%%%%%%%%%%%%%%%%%%%%%%%%%%%%%%%%%%%%%%%%%%%%%%%%%%%%%%%%%%%%%%%%%%%%%%%%%%%%%%%%%%%%%%%%%%%%%%%%%%%%%%%%%%%%%%%%%%%

\section{Preliminaries, Notations, and Backgrounds}\label{sect: backgrounds}

\subsection{Graph}

A graph is denoted as $G = (V, E)$, with $V$ the set of nodes and $E$ the set of edges.
An edge connecting nodes $u_0$ and $u_1$ is denoted as $(u_0, u_1)$. At the same time, $(u_0, u_1)$ is called incident to $u_0$ and/or $u_1$, as well as is regarded as an (incident) edge of $u_0$ and/or $u_1$.
A path with length-$L$ connecting nodes $u_0$ and $u_L$ is denoted as $u_0 u_1 \cdots u_L$, where $u_i \neq u_j$ and $u_j \neq u_t$ for $0 \leq i < j < t \leq L$.
Specially, when $u_0 = u_L$, it forms a length-$L$ (size-$L$) cycle.
$\forall x, y \in V$, if there exists at least one path connecting them, $x$ and $y$ are said to be connected.
In such case, the distance between $x$ and $y$ is defined as the length of the shortest path connecting them.
However, when $x$ and $y$ are not connected, their distance is defined as $\infty$.

\subsection{Binary LDPC Code and Its Tanner Graph}

A binary LDPC code can be represented by its check matrix $\mathbf{H} = [h_{i, j}]_{m \times n}$, where $h_{i, j} \in GF(2)$ for $0 \leq i < m$ and $0 \leq j < n$.
Also, it can be represented by its TG $G = (V_c \cup V_v, E)$, where $V_c = \big\{c_i \big| 0 \leq i < m\big\}$ is the set of the CNs while $c_i$ is the $i$-th CN of the TG which corresponds to the $i$-th row of $\mathbf{H}$, and $V_v = \big\{v_j \big| 0 \leq j < n\big\}$ is the set of the VNs while $v_j$ is the $j$-th VN of the TG which corresponds to the $j$-th column of $\mathbf{H}$, and $(c_i, v_j) \in E$ if and only if $h_{i, j} = 1$.
In this paper, it makes no difference when referring to $\mathbf{H}$ and/or its corresponding TG $G$.
In addition, only $E$ may denote a different set of edges during the construction process of the TG while $V = V_c \cup V_v$ keeps invariant.
Furthermore, denote $E_{v_j} = \big\{(c_i, v_j) \big| 0 \leq i < m\big\}$ for $0 \leq j < n$ as the set including all the incident edges of the VN $v_j$,
and denote $D = \big\{d_{v_j} \big| 0 \leq j < n\big\}$ as the set including the degrees of all VNs, where $d_{v_j}$ is the degree of the VN $v_j$.

\subsection{ACE and ACE Spectrum}

In the TG, the ACE \cite{	Tian04} value of a path is defined as $\sum_j{\big(d_{v_j} - 2\big)}$, where $d_{v_j}$ is the degree of the $j$-th VN of the path, and the summation is taken over all VNs of the path.
The minimum path ACE metric \cite{	Vukobratovic08} between two arbitrary nodes $x$ and $y$ is defined as the minimum ACE value of the shortest paths connecting nodes $x$ and $y$. If nodes $x$ and $y$ are not connected, their minimum path ACE metric is defined as $\infty$.
The ACE spectrum \cite{	Vukobratovic07} of depth $d_{max}$ of a TG $G$ is defined as a $d_{max}$-tuple $\boldsymbol{\eta}(G) = (\eta_2, \eta_4, \ldots, \eta_{2d_{max}})$, where $\eta_{2i},~1 \leq i \leq d_{max}$ is the minimum ACE value of all $2i$-cycles in $G$. If $G$ does not contain any $2i$-cycles, $\eta_{2i}$ is set as $\infty$. Furthermore, the comparison rule between two ACE spectra of depth $d_{max}$ in  \cite{	Vukobratovic08} is defined as
%$\boldsymbol{\eta}^{(1)} > \boldsymbol{\eta}^{(2)} \!\!\iff\!\! \exists j \leq d_{max} \big| \big(\eta_{2j}^{(1)} > \eta_{2j}^{(2)}~\text{and}~\eta_{2i}^{(1)} = \eta_{2i}^{(2)},~1 \leq i < j\big)$.
$\boldsymbol{\eta}^{(1)} > \boldsymbol{\eta}^{(2)} \iff \exists j \leq d_{max} \Big| \Big( \eta_{2j}^{(1)} > \eta_{2j}^{(2)} ~\text{and}~ \eta_{2i}^{(1)} = \eta_{2i}^{(2)},~1 \leq i < j\Big)\text{.}$
In such case, the TG with a larger ACE spectrum  is generally considered to be better \cite{	Vukobratovic07,	Vukobratovic08,	Vukobratovic09, Asvadi12}.

\subsection{Other Notations and Definitions}

In the rest, the following notations are used. Denote:
\begin{itemize}
%\item $m$ as the number of CNs;
%\item $n$ as the number of VNs;
%\item $V_c = \big\{c_i \big| \forall i, 0 \leq i < m\big\}$ as the set of all CNs, where $c_i$ is the $i$-th CN;
%\item $V_v = \big\{v_j \big| \forall j, 0 \leq j < n\big\}$ as the set of all VNs, where $v_j$ is the $j$-th VN;
%\item $E_{v_j} = \big\{(c_i, v_j) \big| \forall i, 0 \leq i < m\big\}$ for $0 \leq j < n$ as the set of all possible edges of the VN $v_j$;
%\item $D = \big\{d_{v_j} \big| \forall j, 0 \leq j < n\big\}$ as the set of degrees of all VNs, where $d_{v_j}$ is the degree of $v_j$;
%\item $G = (V, E)$ as the TG, where $V = V_c \cup V_v$ always holds while $E$ may denote a different set of edges;
%\item $\mathbf{H} = [h_{i, j}]_{m \times n}$ as the parity-check matrix corresponding to $G$; (In this paper, it makes no difference when referring to $G$ and/or $\mathbf{H}$, i.e., $G \triangleq \mathbf{H}$.)
\item $f_{x, y}^{(G)}$, $\forall x, y \in V$ as the \emph{metric} between nodes $x$ and $y$ under $G$. In this paper, each time of using $f_{x, y}^{(G)}$ always refers to one of the following two metrics:
\begin{align}
f_{x, y}^{(G)} &= d_{x, y}^{(G)}\text{,}\label{eqn: metric as distance}\\
f_{x, y}^{(G)} &= \big(d_{x, y}^{(G)}, a_{x, y}^{(G)}\big)\text{,}\label{eqn: metric as dist and ACE pair}
\end{align}
where $d_{x, y}^{(G)}$ indicates the distance and $a_{x, y}^{(G)}$ indicates the minimum path ACE metric  between nodes $x$ and $y$ under $G$ respectively;
\item $F_{X, y}^{(G)} = \big\{f_{x, y}^{(G)} \big| \forall x \in X\big\}, \forall X \subseteq V, \forall y \in V$;
\item $F_{X, Y}^{(G)} = \big\{f_{x, y}^{(G)} \big| \forall x \in X, \forall y \in Y\big\}, \forall X, Y \subseteq V$;
\item $\boldsymbol{0} = 0, \boldsymbol{1} = 1, \boldsymbol{\infty} = \infty, \boldsymbol{0}_{v_j} = 0, \boldsymbol{1}_{v_j} = 1$ if the metric is defined as (\ref{eqn: metric as distance}), otherwise $\boldsymbol{0} = (0, 0), \boldsymbol{1} = (1, 0), \boldsymbol{\infty} = (\infty, \infty), \boldsymbol{0}_{v_j} = (0, d_{v_j} - 2), \boldsymbol{1}_{v_j} = (1, d_{v_j} - 2)$ if the metric is defined as (\ref{eqn: metric as dist and ACE pair}), where $v_j, 0 \leq j < n$ is the $j$-th VN of the TG;
\item $g^{(G)}$ as the \emph{girth} of $G$, indicating the metric of the minimum cycle in $G$; (When the metric of any specific cycle is referred to, if there no corresponding cycle exists, the metric is regarded as $\boldsymbol{\infty}$.)
\item $g_x^{(G)}$, $\forall x \in V$ as the \emph{local girth} of node $x$ under $G$, indicating the metric of the minimum cycle in $G$ which passes through $x$;
\item $g_{(x, y)}^{(G)}$, $\forall x \in V_c, \forall y \in V_v$ as the local girth of edge $(x, y)$ under $G$, indicating the metric of the minimum cycle in $G$ which passes through $(x, y)$;
\item $|X|$ as the cardinality of an arbitrary set $X$;
\item $N$ as the one-dimension circulant-size of $\mathbf{H}$; (I.e., each circulant of $\mathbf{H}$ has size $N \times N$.)
\item $\big\{(c_i, v_j)_N\big\}$ = $\big\{ \big(c_{\pi(i, N, t)},~v_{\pi(j, N, t)} \big) \big| t = 0, 1, \ldots, N - 1 \big\}$ for $0 \leq i < m$ and $0 \leq j < n$, where $\pi(x, N, t) = \lfloor x / N \rfloor \cdot N +\!\!\!\! \mod\!\!(x + t, N)$, while $\lfloor x/N \rfloor$ is the floor of $x/N$ and \!\!\!\!$\mod\!\!(x + t, N)$ is the remainder of $(x + t)$ modulo $N$.
\end{itemize}

In addition, in this paper, the comparison rule between two pairs of (\ref{eqn: metric as dist and ACE pair}) is defined as
$f_{x_0, y_0}^{(G)} > f_{x_1, y_1}^{(G)} \iff
d_{x_0, y_0}^{(G)} > d_{x_1, y_1}^{(G)}~\text{or}~\Big( d_{x_0, y_0}^{(G)} = d_{x_1, y_1}^{(G)} ~\text{and}~ a_{x_0, y_0}^{(G)} > a_{x_1, y_1}^{(G)}\Big)\text{.}$
The addition/subtraction between two pairs of (\ref{eqn: metric as dist and ACE pair}) is defined as
%\begin{displaymath}
$f_{x_0, y_0}^{(G)} \pm f_{x_1, y_1}^{(G)} = \left( d_{x_0, y_0}^{(G)} \pm d_{x_1, y_1}^{(G)}, a_{x_0, y_0}^{(G)} \pm a_{x_1, y_1}^{(G)}\right)\text{.}$
%\end{displaymath}
Furthermore, to measure the VN-local-girth distribution (VNLGD) of a TG under the metric (\ref{eqn: metric as distance}), a polynomial
%\begin{equation*}\label{eqn: VNLGD}
$\phi(x) = \sum \limits_{i > 0} p_ix^i$
%\end{equation*}
is defined, where $p_i$ has indicated the percentage of the VNs whose local girths are $i$ among all the VNs.
If there are no VNs whose local girths are $i$, $p_i$ is considered as $0$ and $p_ix^i$ will be omitted from $\phi(x)$.
Meanwhile, the comparison rule between two VNLGDs is defined as
$\phi^{(1)}(x) < \phi^{(2)}(x) \iff
\exists j > 0 \Big| \Big( p_j^{(1)} < p_j^{(2)} ~\text{and}~ p_i^{(1)} = p_i^{(2)},~i < j\Big)\text{.}$
In such case, the TG with a smaller VNLGD is considered to be better with regard to that it generally contains less small (may be the smallest) cycles.
Finally, an operation $\uplus$, which works between a TG $G$ and an edge $e$ or an edge-set $\tilde{E}$, is defined as
$G \uplus e = \big(V, E \cup \{e\}\big)$ or
$G \uplus \tilde{E} = \big(V, E \cup \tilde{E}\big)$\text{.}

\subsection{PEG Algorithm and ACE Constrained PEG Algorithm}

The PEG algorithm \cite{	Hu05} constructs the TG by establishing edges between VNs and CNs in stages under the metric (\ref{eqn: metric as distance}), where in each stage only one edge is established.
The edges are established in the order from small to large based on the indices of the VNs they are incident to.
%, i.e., $v_0, v_1, \ldots, v_{n-1}$.
To be specific, $v_0$ takes the first $d_{v_0}$ consecutive stages to establish its edges, and then $v_1$ takes the next $d_{v_1}$ consecutive stages to establish its edges, and the construction process continues until $v_{n-1}$ takes the last $d_{v_{n-1}}$ stages to establish its edges.
%In general, the degrees of the aforementioned VN-sequence are nondecreasing, i.e., $d_{v_0} \leq d_{v_1} \leq \cdots \leq d_{v_{n-1}}$.
In such case, the VN, which is of interest to any stage, will be known before starting the construction process.
Then, the VN, which is of interest to the current stage, is conveniently denoted as $v_c$.
Furthermore, by saying the $k$-th stage of $v_c$, where $1 \leq k \leq d_{v_c}$ always holds and sometimes will be omitted in the rest of this paper, we refer to the stage in which the $k$-th edge of $v_c$ is established. %$\mathcal{S}_k^{(v_c)}$ denote the stage via math notation

In each stage of $v_c$, a CN, say $c_i$, is selected based on the \emph{selection strategy} at the beginning of this stage, and then $(c_i, v_c)$ is established at the end of this stage.
The selection strategy is a list of selection criteria of decreasing priority \cite{	Vukobratovic08}. If multiple CNs survive any selection criterion, the next selection criterion is applied on the surviving CNs only.
Otherwise, the single survivor is selected and the selection procedure is terminated.
Let $G$ be the realtime TG setting, then the selection strategy of the PEG algorithm \cite{Hu05} is summarized in the following Strategy \ref{strategy: PEG}, while the pseudo-code of the PEG algorithm \cite{Hu05} is presented in Algorithm \ref{algo: PEG}.

\begin{strategy}[Selection strategy of the PEG algorithm \cite{	Hu05}]
\label{strategy: PEG}
\end{strategy}
\begin{enumerate}
\item Select the CN $c_i$ that $f_{c_i, v_c}^{(G)} = \max \limits_{0 \leq j < m}f_{c_j, v_c}^{(G)}\text{;}$
%\begin{equation*}
%f_{c_i, v_c}^{(G)} = \max \limits_{0 \leq j < m}f_{c_j, v_c}^{(G)}\text{;}
%\end{equation*}
\item Select the survivor whose degree is minimal;\label{criterion: strategy PEG @ minimal degree}
\item Select the survivor randomly.\label{criterion: strategy PEG @ random}
\end{enumerate}

\begin{algorithm}[h]
\caption{The PEG Algorithm \cite{Hu05}}
\label{algo: PEG}
\begin{algorithmic}[1]
\REQUIRE $m, n, D$.
%\ENSURE $\mathbf{H}$.
%\STATE $\mathbf{H} = m \times n$ zero matrix.\label{code: algo PEG @ E = empty}
\ENSURE $G$.
%\STATE $G = (V, E) = (V_c \cup V_v, \emptyset)$.\label{code: algo PEG @ E = empty}
\STATE $G = (V, \emptyset)$.\label{code: algo PEG @ E = empty}
\FOR {$j = 0;~j < n;~j+\!+$}\label{code: algo PEG @ for v_j}
%    \STATE //\textit{$v_j$ is the VN current in interest, i.e., $v_c = v_j$.}
    \STATE $f_{c_i, v_j}^{(G)} = \boldsymbol{\infty}$ for $0 \leq i < m$.
    \FOR {$k = 1;~k \leq d_{v_j};~k ++$}\label{code: algo PEG @ for k-th edge}
        \STATE $c_i =$ the CN selected based on Strategy \ref{strategy: PEG}.\label{code: algo PEG @ selection}
        %\STATE $h_{i, j} = 1$.
        %\STATE $E = E \cup (c_i, v_j)$.
        \STATE $G = G \uplus (c_i, v_j)$.
        \STATE Calculate $F_{V_c, v_j}^{(G)}$.\label{code: algo PEG @ calc metrics}
    \ENDFOR\label{code: algo PEG @ end for k}
\ENDFOR\label{code: algo PEG @ end for v_j}
\RETURN $G$.
\end{algorithmic}
\end{algorithm}

\begin{remark}
Assuming that $c_i$ is selected based on Strategy \ref{strategy: PEG}, according  to Corollary \ref{coro: easy calculation of g(vc, 1)} (refer to Section \ref{subsect: Calculate the Multi-Edge Local Girths}), it holds that
$g_{v_c}^{(G \uplus (c_i, v_c))} = g_{(c_i, v_c)}^{(G \uplus (c_i, v_c))} = f_{c_i, v_c}^{(G)} + 1\text{,}$
implying that the establishment of $(c_i, v_c)$ makes $v_c$ achieve its maximum realtime local girth.
 %which is measured after one edge in $E_{v_c} \setminus E$ will have been established.
\end{remark}

\begin{remark}
In Algorithm \ref{algo: PEG}, line \ref{code: algo PEG @ calc metrics} is the most time-consuming part. Hu \textit{et al.} \cite{Hu05} employed a breadth-first search (BFS) \cite{	introAlgo01} like method to implement this calculation.
In the worst case, the computational complexity of implementing the BFS once in $G$ is $O\big(m + |E|\big)$. As the calculation is totally implemented $|E|$ times, the total computational complexity of the original PEG algorithm \cite{Hu05} thus is $O\big(m|E| + |E|^2\big)$.
\end{remark}

The key idea of the ACE constrained PEG algorithm\footnote{The ACE constrained PEG algorithm introduced in this paper is a little different from the original one in \cite{	Vukobratovic08}. See \cite{	Vukobratovic08} for more details.} \cite{	Vukobratovic08} is to construct the LDPC codes under the framework of the PEG algorithm \cite{Hu05}, (Refer to Strategy \ref{strategy: PEG} and Algorithm \ref{algo: PEG}.) while replacing the metric (\ref{eqn: metric as distance}) with the metric (\ref{eqn: metric as dist and ACE pair}).
However, the modification on the metric makes the ACE constrained PEG algorithm \cite{	Vukobratovic08} a better algorithm, compared to  the PEG algorithm \cite{Hu05}, the ACE algorithm \cite{	Tian04}, \textit{etc}, in terms of designing the LDPC codes with better ACE spectra and better error performances.
In addition, the total computational complexity of the ACE constrained PEG algorithm \cite{	Vukobratovic08} remains the same as that of the PEG algorithm \cite{Hu05}, i.e., $O\big(m|E| + |E|^2\big)$.

\subsection{Circulant, CPM, and QC-LDPC Code}

A circulant is a square matrix where each row-vector is cyclically shifted one element to the right relative to the preceding row-vector. (The first row-vector's preceding row-vector is the last row-vector.) Consequently, a zero square matrix is also regarded as a circulant, i.e., the zero circulant. If there is only one entry of 1 in each row and each column of a circulant and 0s elsewhere, the circulant is considered as a CPM.

If the check matrix of an LDPC code is an array of sparse circulants with the same size, then it is a QC-LDPC code. The check matrix of a QC-LDPC code typically looks like follows:
\begin{equation}
\label{eqn: QC check matrix}
\mathbf{H} =
\begin{bmatrix}
\mathbf{H}_{0, 0} & \mathbf{H}_{0, 1} & \cdots & \mathbf{H}_{0, K-1}\\
\mathbf{H}_{1, 0} & \mathbf{H}_{1, 1} & \cdots & \mathbf{H}_{1, K-1}\\
\vdots & \vdots & \ddots & \vdots\\
\mathbf{H}_{J-1, 0} & \mathbf{H}_{J-1, 1} & \cdots & \mathbf{H}_{J-1, K-1}\\
\end{bmatrix}\text{,}
\end{equation}
where $\mathbf{H}_{i, j},~0 \leq i < J,~0 \leq j < K$ are circulants of the same size.
Furthermore, in the TG $G$ of a QC-LDPC code, for $0 \leq i < m, 0 \leq j < n$, it obviously holds that
\begin{equation}
\label{eqn: single edge corresponds to cyclic edges}
(c_i, v_j) \in E \iff \big\{(c_i, v_j)_N\big\} \subseteq E\text{.}
\end{equation}
%In addition, $\forall e \in \big\{(c_i, v_j)_N\big\}$, it holds that
%\begin{equation}
%\label{eqn: local girths of cyclic edges equal}
%g_{(c_i, v_j)}^{(G)} = g_e^{(G)}\text{,}
%\end{equation}
%which can be easily proved based on (\ref{eqn: single edge corresponds to cyclic edges}).

\subsection{QC-PEG Algorithm}

The QC-PEGA \cite{	Li04} constructs the QC-LDPC codes with any predefined valid design parameters. Li \emph{et al.} \cite{Li04} implemented the QC-PEGA similar to  the PEG algorithm \cite{Hu05}, where both the metric and the selection strategy remain the same.
However, because of (\ref{eqn: single edge corresponds to cyclic edges}), the QC-PEGA \cite{Li04} cyclically establishes $N$ edges in a single circulant at a time in each stage, instead of only one edge in each stage as that in the PEG algorithm \cite{Hu05}.
To be specific, the pseudo-code of the QC-PEGA \cite{Li04} is presented in Algorithm \ref{algo: QC-PEG}.

\begin{algorithm}[h]
\caption{The QC-PEG Algorithm \cite{Li04}}
\label{algo: QC-PEG}
\begin{algorithmic}[1]
\REQUIRE $m, n, N, D$.
%\ENSURE $\mathbf{H}$.
%\STATE $\mathbf{H} = m \times n$ zero matrix.
\ENSURE $G$.
%\STATE $G = (V, E) = (V_c \cup V_v, \emptyset)$.
\STATE $G = (V, \emptyset)$.
\FOR {$j = 0;~j < n;~j\,+\!\!= N$}
%    \STATE //\textit{$v_j$ is the VN current in interest, i.e., $v_c = v_j$.}
    \STATE $f_{c_i, v_j}^{(G)} = \boldsymbol{\infty}$ for $0 \leq i < m$.
    \FOR {$k = 1;~k \leq d_{v_j};~k ++$}
        \STATE $c_i =$ the CN selected based on Strategy \ref{strategy: PEG}.
        %\STATE $\mathbf{H}_{\lfloor i / N \rfloor, j/N} = \mathbf{H}_{\lfloor i / N \rfloor, j/N} + \mathbf{P}_N^{\mod(i, N)}$.\label{code: algo QC-PEG @ estabish N edges}
        %\STATE $E = E \cup \big\{(c_i, v_j)_N\big\}$.\label{code: algo QC-PEG @ estabish N edges}
        \STATE $G = G \uplus \big\{(c_i, v_j)_N\big\}$.\label{code: algo QC-PEG @ estabish N edges}
        \STATE Calculate $F_{V_c, v_j}^{(G)}$.\label{code: algo QC-PEG @ calc metrics}
    \ENDFOR
\ENDFOR
\RETURN $G$.
\end{algorithmic}
\end{algorithm}

\begin{remark}\label{rmk: valid parameters}
Algorithm \ref{algo: QC-PEG}  generates the check matrix of a QC-LDPC code with the form (\ref{eqn: QC check matrix}), where $J = m/N$, $K = n/N$, and each circulant has size $N \times N$. Therefore, input parameters (design parameters) are considered valid if and only if: 1) $m$ and $n$ are multiples of $N$. 2) $0 < d_{v_j} = d_{v_{j+1}} = \cdots = d_{v_{j+N-1}} \leq m$ for $j = 0, N, \ldots, n - N$.
\end{remark}

%\begin{remark}
%For $j = 0, N, \ldots, n - N$, since there exists a one-to-one mapping between $F_{V_c, v_j}^{(G)}$ and $F_{V_c, v_t}^{(G)}$ s.t. $\lfloor j/N \rfloor = \lfloor t/N \rfloor$, which can be easily proved based on (\ref{eqn: local girths of cyclic edges equal}), that only the metrics in $F_{V_c, v_j}^{(G)}$ are considered by Algorithm \ref{algo: QC-PEG}.
%\end{remark}

\begin{figure}[!t]
\centering
\includegraphics[scale = 0.5]{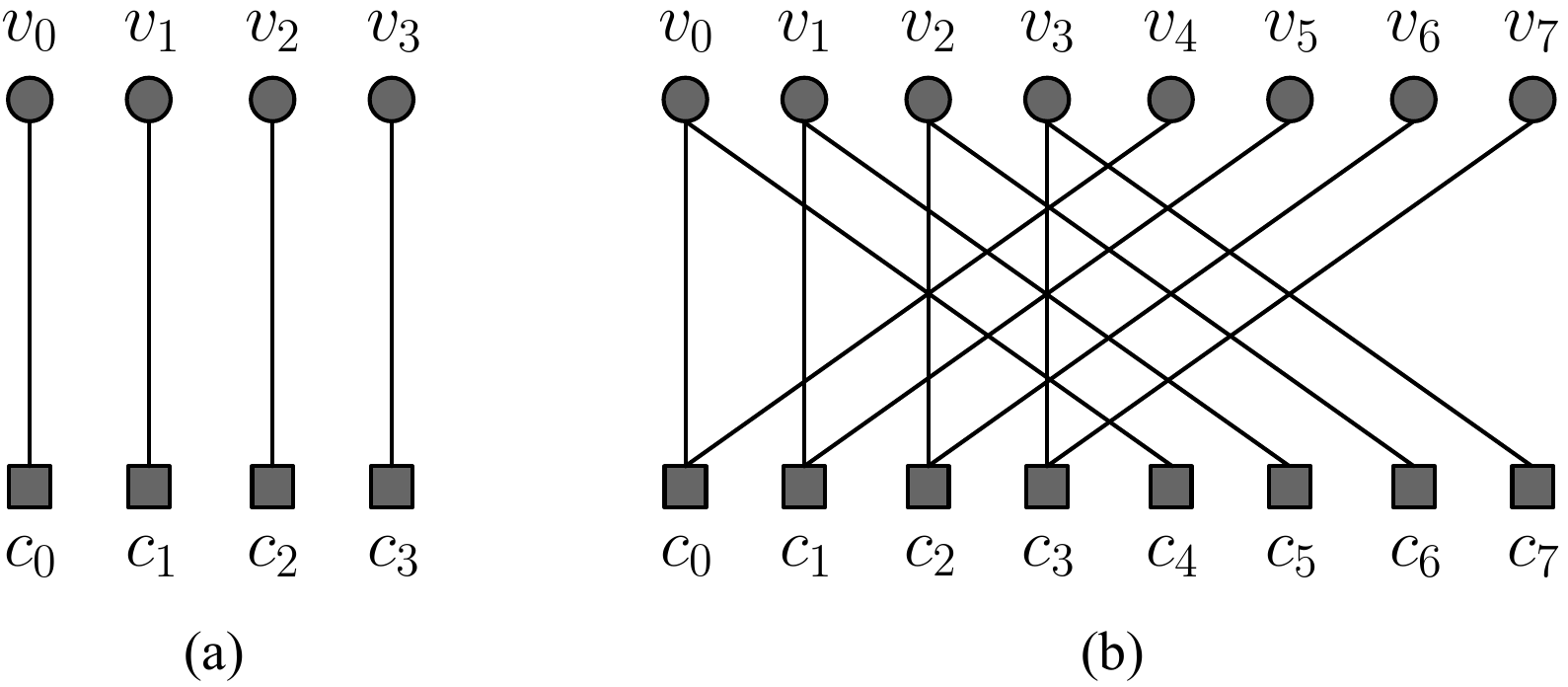}
\vspace{-3mm}
\caption{Undetectable shortest cycles resulted by the QC-PEGA \cite{	Li04} and the CP-PEGA \cite{	Lin08}. (a) Undetectable 4-cycles resulted by the QC-PEGA \cite{	Li04}. (b) Undetectable 8-cycles resulted by the CP-PEGA \cite{	Lin08}.}
\label{fig: undetectableCycles}
\vspace{-3mm}
\end{figure}

\begin{remark}\label{rmk: undetectable cycles}
In line \ref{code: algo QC-PEG @ estabish N edges} of Algorithm \ref{algo: QC-PEG}, $N$ edges are established in a single circulant at a time.
But the QC-PEGA \cite{	Li04} cannot real timely detect the cycles which contain two or more newly established edges. As a result, the QC-PEGA \cite{	Li04} sometimes results in 4-cycles just in a single circulant of the check matrix when there contain multiple edges.
A typical example is given in Fig. \ref{fig: undetectableCycles}(a), where $c_2$ has the chance to survive Strategy \ref{strategy: PEG} as well as to be selected for establishing the second edge of $v_0$, and 4-cycles will form in the single circulant if edges in $\big\{(c_2, v_0)_4\big\}$ are established.
To avoid generating 4-cycles in a single circulant of the check matrix, the CP-PEGA \cite{	Lin08} requires each nonzero circulant of the check matrix to be a CPM during the construction process of the QC-PEGA \cite{Li04}.
However, this modification cannot real timely detect the cycles which contain two or more newly established edges either. Fig. \ref{fig: undetectableCycles}(b) shows an example of such case, where $c_6$ has the chance to survive Strategy \ref{strategy: PEG} as well as to be selected for establishing the second edge of $v_4$, and 8-cycles will form if edges in $\big\{(c_6, v_4)_4\big\}$ are established.
In addition, it can be easily proved that $8$ is the shortest length of the undetectable cycles, which contain two or more newly established edges, in the QC-LDPC code designed by the CP-PEGA \cite{	Lin08}.
Furthermore, because of the CPM-requirement, the CP-PEGA \cite{	Lin08} additionally requires the maximum CN-degree not to exceed $m / N$ (the number of circulants in a column), or it will fail the construction.
%\begin{equation}
%\label{eqn: 4-cycle circulant}
%\begin{bmatrix}
%1 & 0 & 1 & 0\\
%0 & 1 & 0 & 1\\
%1 & 0 & 1 & 0\\
%0 & 1 & 0 & 1\\
%\end{bmatrix}
%\end{equation}
\end{remark}

\begin{remark}
The construction process of Algorithm \ref{algo: QC-PEG} is accelerated by a factor of $1/N$ compared to that of Algorithm \ref{algo: PEG}. Thus, the total computational complexity of the QC-PEGA \cite{Li04} is $O\big((m|E| + |E|^2) / N\big)$.
\end{remark}

%%%%%%%%%%%%%%%%%%%%%%%%%%%%%%%%%%%%%%%%%%%%%%%%%%%%%%%%%%%%%%%%%%%%%%%%%%%%%%%%%%%%%%%%%%%%%%%%%%%%%%%%%%%%%%%%%%%%%%%%%%%%%%%%%%%%%

\section{Multi-Edge Metric-Constrained PEG Algorithm}\label{sect: MM-PEGA}

\subsection{MM-PEGA}

Since the PEG algorithm \cite{Hu05} and the ACE constrained PEG algorithm \cite{	Vukobratovic08} only differ at their metrics, this paper unifies them to one integrated algorithm, called the M-PEGA, where the PEG algorithm \cite{	Hu05} is referred to when the metric is defined as (\ref{eqn: metric as distance}), and the ACE constrained PEG algorithm \cite{	Vukobratovic08} is referred to when the metric is defined as (\ref{eqn: metric as dist and ACE pair}).
As in each stage of the M-PEGA, a CN is selected to primarily maximize the local girth of $v_c$ whenever a new edge is added to $v_c$.
In such case, the final local girth of $v_c$, which is measured at the end of the $d_{v_c}$-th stage of $v_c$, may not be optimal.
This situation is demonstrated in the following example.

\begin{figure}[!t]
\centering
\includegraphics[scale = 0.5]{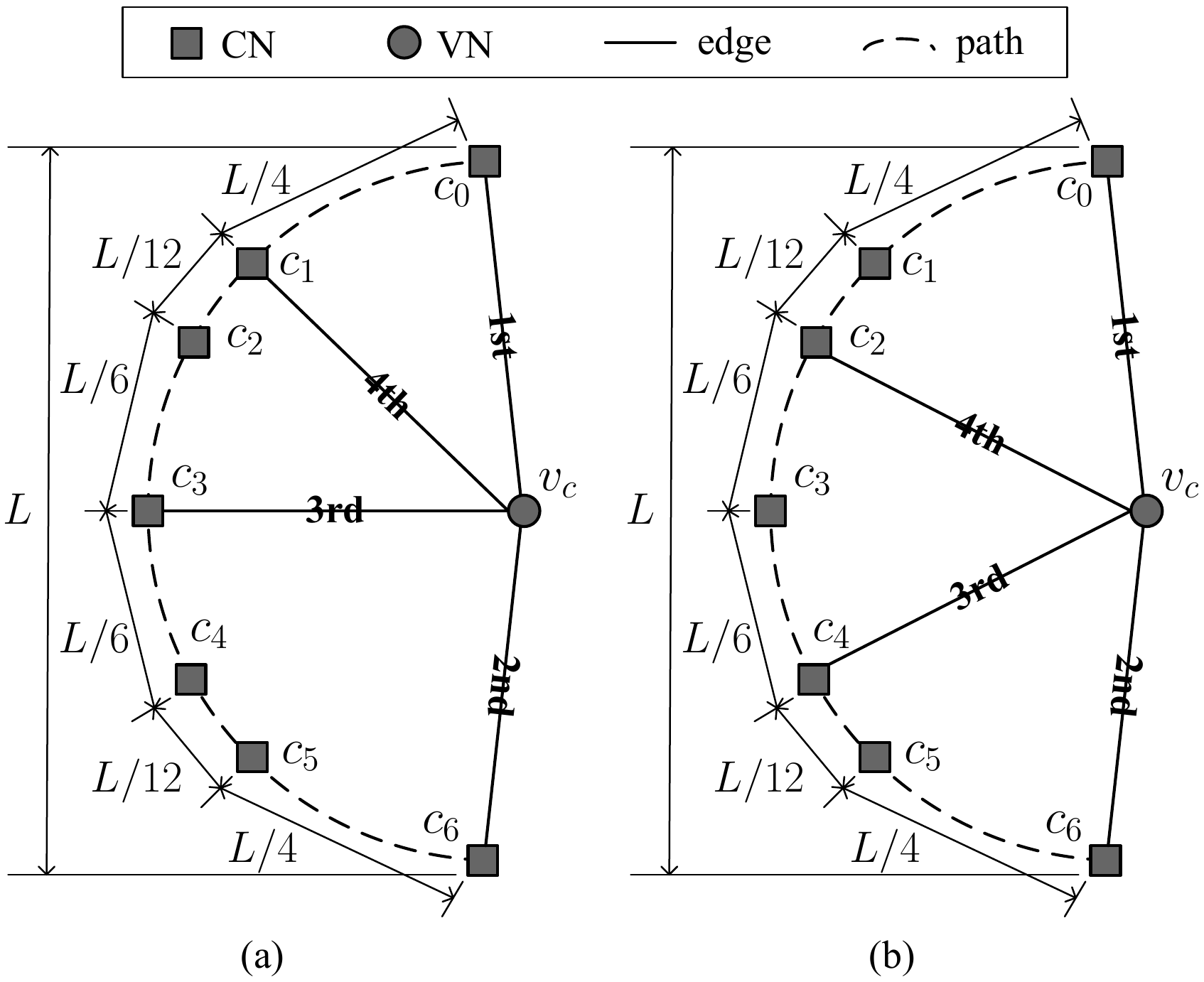}
\vspace{-3mm}
\caption{The final local girth of $v_c$ under the metric (\ref{eqn: metric as distance}) achieved by the M-PEGA \textit{vs} that achieved by the two-edge M-PEGA. (a) Edges are established by the M-PEGA. (b) Edges are established by the two-edge M-PEGA.}
\label{fig: localGirths}
\vspace{-3mm}
\end{figure}

\begin{example}
\label{eg: MCPEG vs MSMCPEG @ local girth}
Fig. \ref{fig: localGirths} presents two simple TGs. At the beginning, each of the TGs consists of the VN $v_c$ and the path connecting CNs $c_0$ and $c_6$. Then, four edges are established between $v_c$ and some CNs in the path by some specific design algorithm.
In addition, only CNs $c_0$--$c_6$ are displayed in the path of each TG, while the other CNs are omitted since they have no chances to be selected for establishing the edges of $v_c$ under the considered design algorithms.
Assume that the metric is defined as (\ref{eqn: metric as distance}), and then denote the lengths (distances) between different pairs of CNs in each path as expressions of $L$.
As Fig. \ref{fig: localGirths} aims to compare the relative sizes between the final local girths of $v_c$ under different design algorithms, only assume that $L$ is a valid positive integer while its value is not in the interest.
Fig. \ref{fig: localGirths}(a) shows how the M-PEGA establishes the four edges of $v_c$, where the edges are labeled 1st, 2nd, 3rd, and 4th based on the order of being established.
According to Strategy \ref{strategy: PEG}, in Fig. \ref{fig: localGirths}(a), the local girth of $v_c$ has been maximized whenever a new edge is added to $v_c$.
In contrast, another case is shown in Fig. \ref{fig: localGirths}(b). (The construction process will be shown in detail in Example \ref{eg: two-edge M-PEGA constructs TG}.)
The establishment of the third edge in Fig. \ref{fig: localGirths}(b) is considered to be worse than that in Fig. \ref{fig: localGirths}(a) in terms of making the realtime local girth of $v_c$ maximum, where $g_{v_c}^{(G_b)} = L/3 + 2$ in Fig. \ref{fig: localGirths}(b) \textit{vs} $g_{v_c}^{(G_a)} = L/2 + 2$ in Fig. \ref{fig: localGirths}(a) at the end of the third stage of $v_c$. (Here in this example, $G_a$ and $G_b$ temporarily represent the realtime TG settings in Fig. \ref{fig: localGirths}(a) and in Fig. \ref{fig: localGirths}(b) respectively.)
Nonetheless, if it is known that where the fourth edge will be established at the beginning of the third stage of $v_c$, the establishment of the third edge in Fig. \ref{fig: localGirths}(b) is considered to be better than that in Fig. \ref{fig: localGirths}(a) in terms of making the final local girth of $v_c$ maximum, where $g_{v_c}^{(G_b)} = L/3 + 2$ in Fig. \ref{fig: localGirths}(b) \textit{vs} $g_{v_c}^{(G_a)} = L/4 + 2$ in Fig. \ref{fig: localGirths}(a) at the end of the fourth stage of $v_c$.
\end{example}

Inspired by Example \ref{eg: MCPEG vs MSMCPEG @ local girth}, this paper proposes an improvement for the M-PEGA, called the MM-PEGA, to further optimize the cycle-structure of the non-QC LDPC code.
On the whole, the MM-PEGA is implemented under the framework of Algorithm \ref{algo: PEG} but adopts a different selection strategy.
%On the whole, when establishing an edge of $v_c$ at one stage, instead of considering to maximize the realtime local girth of $v_c$ after one valid edge will have been established as that in the M-PEGA, the MM-PEGA considers to maximize the local girth of $v_c$  after more valid edges will have been established, just like that the establishment of the third edge in Fig. \ref{fig: localGirths}(b) considers to maximize the local girth of $v_c$ after the fourth edge will have been established.
To illustrate the MM-PEGA as well as its selection strategy, the multi-edge local girths are defined firstly,

Assume that $G = (V_c \cup V_v, E)$ is the current TG setting, and let $r$ be a positive integer.
Then, under $G$ and $v_c$,  a length-$r$ \emph{CN-sequence (CNS)} $\mathbf{S} = (s_i)_{1 \leq i \leq r}$, where $s_i \in V_c$ for $1 \leq i \leq r$, is called \emph{valid} if $s_i \neq s_j$ for $1 \leq i < j \leq r$ and $(s_i, v_c) \notin E$ for $1 \leq i \leq r$.
In addition, the corresponding \textit{TG-sequence (TGS)} of $\mathbf{S}$ is defined as $(G_i)_{1 \leq i \leq r+1}$, where $G_1 = G$ and $G_{i+1} = G_i \uplus (s_i, v_c)$ for $1 \leq i \leq r$.

\begin{definition}\label{def: definition of multi-edge local girths}
Under $G$, $v_c$, and $r$, let $\mathbf{S}_{v_c}^{(r, G)}$ be a length-$r$ valid CNS with its corresponding TGS $(G_i^{(\mathbf{S})})_{1 \leq i \leq r+1}$.
Then, the \emph{$r$-edge local girth of $v_c$} is defined as
\begin{equation}
\label{eqn: r-step local girth of vc}
g_{v_c}^{(r, G)} = \max \limits_{\mathbf{S}_{v_c}^{(r, G)}} g_{v_c}^{(G_{r+1}^{(\mathbf{S})})}\text{,}
\end{equation}
and the \emph{$r$-edge local girth of $c_i$}, $0 \leq i < m$ is defined as
\begin{equation}
\label{eqn: r-step local girth of ci}
g_{c_i, v_c}^{(r, G)} =
\begin{cases}
-\boldsymbol{\infty} & \text{if } (c_i, v_c) \in E\text{,}\\
g_{v_c}^{(G \uplus (c_i, v_c))} & \text{else if } r = 1\text{,}\\
g_{v_c}^{(r - 1, G \uplus (c_i, v_c))} & \text{otherwise.}
\end{cases}
\end{equation}
\end{definition}

In (\ref{eqn: r-step local girth of vc}), $g_{v_c}^{(r, G)}$ indicates the maximum local girth of $v_c$  measured after any $r$ edges in $E_{v_c} \setminus E$ will have been established. (All the $r$ edges are incident to $v_c$ while have not been established in $G$.)
Naturally, for some length-$r$ valid CNS $\mathbf{\hat{S}}_{v_c}^{(r, G)}$ with corresponding TGS $(\hat{G}_i)_{1 \leq i \leq r+1}$, if $g_{v_c}^{(r, G)} = g_{v_c}^{(\hat{G}_{r+1})}$, then $\mathbf{\hat{S}}_{v_c}^{(r, G)}$ is called \emph{optimal} for $g_{v_c}^{(r, G)}$.
In (\ref{eqn: r-step local girth of ci}), if $(c_i, v_c) \notin E$, $g_{c_i, v_c}^{(r, G)}$ indicates the maximum local girth of $v_c$ measured  after $(c_i, v_c)$ as well as any $r - 1$ edges in $E_{v_c} \setminus E \setminus \{(c_i, v_c)\}$ will have been established.
It obviously holds that
\begin{equation}
\label{eqn: gvr = max gcr}
g_{v_c}^{(r, G)} = \max \limits_{0 \leq i < m} g_{c_i, v_c}^{(r, G)}\text{\qquad s.t. \quad $r > 0$.}
\end{equation}

\begin{definition}
\label{def: definition of MM-PEGA, edge-trials}
The \emph{edge-trials} is a maximum number of edges tried by a specific MM-PEGA  in each stage to construct a local optimal edge of the TG.
The MM-PEGA with the edge-trials, say  $r$, is called the $r$-edge M-PEGA.
\end{definition}

The edge-trials of a specific MM-PEGA is closely related to its selection strategy.
More precisely, under the current TG setting $G$, the following Strategy \ref{strategy: r-step M-PEGA} is adopted by the $r$-edge M-PEGA to at the beginning of the $k$-th stage of $v_c$, where $k$ is defined in line \ref{code: algo PEG @ for k-th edge} of Algorithm \ref{algo: PEG}, and $r_k$ in Strategy \ref{strategy: r-step M-PEGA} is defined as
\begin{equation}
\label{eqn: r_k = min (r, dvc - k + 1)}
r_k = \min \left\{r, d_{v_c} - k + 1\right\}\text{.}
\end{equation}

\begin{strategy}[Selection strategy of the $r$-edge M-PEGA]
\label{strategy: r-step M-PEGA}
\end{strategy}
\begin{enumerate}
\item Select the CN $c_i$ that $g_{c_i, v_c}^{(r_k, G)} = g_{v_c}^{(r_k, G)}\text{;}$
%\begin{equation*}
%g_{c_i, v_c}^{(r_k, G)} = g_{v_c}^{(r_k, G)}\text{;}
%\end{equation*}
\item Select the survivor $c_i$ that
$g_{(c_i, v_c)}^{(G \uplus (c_i, v_c))} =
\!\max \left\{ g_{(x, v_c)}^{(G \uplus (x, v_c))} \Big|  x \in V_c,  g_{x, v_c}^{(r_k, G)} = g_{v_c}^{(r_k, G)} \right\}\text{;}$
\item Select the survivor whose degree is minimal;
\item Select the survivor randomly.
\end{enumerate}

Since the multi-edge local girth of $v_c$ should be considered under the premise of that $v_c$ only has $d_{v_c}$ edges to be established, (\ref{eqn: r_k = min (r, dvc - k + 1)}) is thus reasonable.
In addition, in Strategy \ref{strategy: r-step M-PEGA}, the first selection criterion selects the CN whose $r_k$-edge local girths is optimal, aiming to make $v_c$ achieve its maximum local girth after $r_k$ new edges are added to $v_c$.
Then, from the CNs which have survived the first selection criterion, the second selection criterion selects a survivor to establish the edge with the largest local girth, aiming to avoid generating cycles with smaller sizes.

\begin{table}[!t]
\renewcommand{\arraystretch}{1.0}
\caption{Two-Edge Local Girths of CNs in Different Stages of the Construction Process in Example \ref{eg: two-edge M-PEGA constructs TG}}
\label{table: two-edge local girths}
\centering
\begin{tabular}{|c|c|c|c|c|}
%\begin{tabular}{|p{0.6cm}<{\centering}|p{1.1cm}<{\centering}|p{1.1cm}<{\centering}|p{1.1cm}<{\centering}|p{1.1cm}<{\centering}|}
%\begin{tabular}{p{0.6cm}<{\centering}p{1.1cm}<{\centering}p{1.1cm}<{\centering}p{1.1cm}<{\centering}p{1.1cm}<{\centering}}
\hline
\bfseries CN    &   \bfseries 1st   &   \bfseries 2nd   &   \bfseries 3rd   &   \bfseries  4th \\
\hline%\hline
$c_0$   &   $L+2$       &   $-\infty$   &   $-\infty$   &   $-\infty$   \\%\hline
$c_1$   &   $3L/4+2$    &   $L/4+2$     &   $L/4+2$     &   $L/4+2$     \\%\hline
$c_2$   &   $2L/3+2$    &   $L/3+2$     &   $L/3+2$     &   $L/3+2$     \\%\hline
$c_3$   &   $L/2+2$     &   $L/2+2$     &   $L/4+2$     &   $L/6+2$     \\%\hline
$c_4$   &   $2L/3+2$    &   $L/3+2$     &   $L/3+2$     &   $-\infty$   \\%\hline
$c_5$   &   $3L/4+2$    &   $3L/8+2$    &   $L/4+2$     &   $L/12+2$    \\%\hline
$c_6$   &   $L+2$       &   $L/2+2$     &   $-\infty$   &   $-\infty$   \\
\hline
\end{tabular}
\vspace{-3mm}
\end{table}

\begin{example}\label{eg: two-edge M-PEGA constructs TG}
This example illustrates how the four edges of $v_c$ in Fig. \ref{fig: localGirths}(b) are established by the two-edge M-PEGA under the metric (\ref{eqn: metric as distance}), where the edges are also labeled 1st, 2nd, 3rd, and 4th based on the order of being established.
The two-edge local girths of CNs $c_0$--$c_6$ measured at the beginning of different stages of $v_c$ are presented in Table \ref{table: two-edge local girths}.
For instance, the column labelled by 1st displays the two-edge local girths of CNs $c_0$--$c_6$ measured at the beginning of  the first stage of $v_c$,
and the column labelled by 2nd displays the two-edge local girths of CNs $c_0$--$c_6$ measured at the beginning of  the second stage of $v_c$, and so on.
However, because of (\ref{eqn: r_k = min (r, dvc - k + 1)}) and Strategy \ref{strategy: r-step M-PEGA}, the last column labelled by 4th displays the one-edge instead of the two-edge local girths of CNs $c_0$--$c_6$ measured at the beginning of  the fourth stage of $v_c$.
To be more specific, the following four stages show how the four edges of $v_c$ are established in Fig. \ref{fig: localGirths}(b) in detail:
1) At the beginning of the first stage of $v_c$, only $c_0$ and $c_6$ survive the first selection criterion of Strategy \ref{strategy: r-step M-PEGA}. Furthermore, $c_0$ and $c_6$ perform the same during the following comparisons until $c_0$ is selected at last by randomness. Then, $(c_0, v_c)$ is established as the first edge of $v_c$.
2) At the beginning of the second stage of $v_c$, only $c_3$ and $c_6$ survive the first selection criterion of Strategy \ref{strategy: r-step M-PEGA}. However, $c_6$ is selected in this stage since it defeats $c_3$ on the second selection criterion of Strategy \ref{strategy: r-step M-PEGA}. Then, $(c_6, v_c)$ is established as the second edge of $v_c$.
3) At the beginning of the third stage of $v_c$, only $c_2$ and $c_4$ survive the first selection criterion of Strategy \ref{strategy: r-step M-PEGA}. Furthermore, $c_2$ and $c_4$ perform the same during the following comparisons until $c_4$ is selected at last by randomness. Then, $(c_4, v_c)$ is established as the third edge of $v_c$.
4) Finally, at the beginning of the fourth stage of $v_c$, only $c_2$ survives the first selection criterion of Strategy \ref{strategy: r-step M-PEGA}. Then, $(c_2, v_c)$ is established as the fourth edge of $v_c$.

\end{example}

Fig. \ref{fig: localGirths} presents a specific case to show that $v_c$ may achieve a larger final local girth under the TG designed by the two-edge M-PEGA, compared to the final local girth achieved under the TG designed by the M-PEGA.
In addition, the design under the two-edge M-PEGA in Fig. \ref{fig: localGirths}(b) is optimal (with regard to making the final local girth of $v_c$ maximum).
Furthermore, any MM-PEGA with an edge-trials larger than one can achieve the optimal design, while the one-edge M-PEGA achieves the design as same as that under the M-PEGA  in Fig. \ref{fig: localGirths}(a).
(It can be seen in Corollary \ref{coro: easy calculation of g(vc, 1)} that the one-edge M-PEGA is equivalent to the M-PEGA.)
However, the MM-PEGA with a larger edge-trials may not always achieve a better design.
For example, change $d_{v_c}$ to 5 in Fig. \ref{fig: localGirths}, then the final local girth of $v_c$ under the two-edge M-PEGA will be $2L/9 + 2$, while any MM-PEGA with an edge-trials except two makes $v_c$ achieve its optimal final local girth $L/4 + 2$.
In general, under the same TG setting at the beginning of the first stage of $v_c$, since the $d_{v_c}$-edge M-PEGA always makes $v_c$ achieve its maximum final local girth, it's expected that increasing the edge-trials of the MM-PEGA will have a positive effect on the final local girth of $v_c$ as well as the cycle-structure of the target TG.
%Meanwhile, it's also expected that increasing the edge-trials of the MM-PEGA will have a positive effect on the cycle-structure as well as the error performance of the LDPC code designed by the MM-PEGA.

\subsection{Calculate the Multi-Edge Local Girths}\label{subsect: Calculate the Multi-Edge Local Girths}

Since the $r$-edge M-PEGA changes the Strategy \ref{strategy: PEG} used in line \ref{code: algo PEG @ selection} of Algorithm \ref{algo: PEG} to the Strategy \ref{strategy: r-step M-PEGA}, the multi-edge local girth of $v_c$ as well as that of all the CNs need to be calculated between line \ref{code: algo PEG @ for k-th edge} and line \ref{code: algo PEG @ selection} of Algorithm \ref{algo: PEG}, right before each time of applying Strategy \ref{strategy: r-step M-PEGA}.
Before proposing the specific algorithm for calculating the multi-edge local girths, we would like to propose some propositions and corollaries to discuss the properties of the multi-edge local girths firstly.

%\begin{definition}
%\label{def: definition of selected CNs, alpha, beta, etc}
Assume that the TG has been constructed by the $r$-edge M-PEGA.
For $1 \leq k \leq d_{v_c}$, denote $\hat{c}_k$ as the CN selected in the $k$-th stage of $v_c$, and denote $(\hat{G}_i)_{1 \leq i \leq d_{v_c} + 1}$ as the corresponding TGS of $(\hat{c}_i)_{1 \leq i \leq d_{v_c}}$, where $\hat{G}_1$ denotes the TG that  at the beginning of the first stage of $v_c$,
Then, under $\hat{G}_k$,  denote $\mathbf{S}^{(k)} = \big(s_i^{(k)}\big)_{1 \leq i \leq r_k}$ as an arbitrary length-$r_k$ valid CNS with its corresponding TGS $\big( G_i^{(k)} \big)_{1 \leq i \leq r_k + 1}$, where $G_1^{(k)} = \hat{G}_k$ obviously holds.
Finally, define
$\alpha_k = \min \limits_{1 \leq i < k} g_{(\hat{c}_i, v_c)}^{(\hat{G}_{i + 1})}\label{eqn: definition of alpha}$ and
$\beta_k = \min \limits_{1 \leq i \leq r_k} g_{(s_i^{(k)}, v_c)}^{(G_{i + 1}^{(k)})}\label{eqn: definition of beta}\text{.}$
%where $\alpha_k$ indicates the realtime local girth of $v_c$ measured at the beginning of the $k$-th stage of $v_c$.
%\end{definition}

%%%%%%%%%%%%%%%% proposition 1 %%%%%%%%%%%%%%%%%%%%%%%

\begin{proposition}
\label{prop: gvc > beta}
For $1 \leq k \leq d_{v_c}$, it holds that
$g_{v_c}^{(r_k, \hat{G}_k)}
= g_{\hat{c}_k, v_c}^{(r_k, \hat{G}_k)}
\geq g_{v_c}^{(G_{r_k + 1}^{(k)})}
= \min \left\{ \alpha_k, \beta_k \right\} = \beta_k\label{eqn: gvc > beta}\text{.}$
\end{proposition}

\begin{IEEEproof}
For $1 \leq k \leq d_{v_c}$, it obviously holds that
$g_{v_c}^{(r_k, \hat{G}_k)} = g_{\hat{c}_k, v_c}^{(r_k, \hat{G}_k)} \geq g_{v_c}^{(G_{r_k + 1}^{(k)})} = \min \left\{ \alpha_k, \beta_k \right\}\text{.}$
In addition, for $k = 1$, $\min \left\{ \alpha_k, \beta_k \right\} = \beta_k$ obviously holds.
Instead, for $2 \leq k \leq d_{v_c}$, by assuming that $\alpha_k < \beta_k$, a contradiction is found in the following proof.

According to the assumption that $\alpha_k < \beta_k$, there exists some index $i, 1 \leq i < k$ satisfying
$\alpha_k
= g_{(\hat{c}_i, v_c)}^{(\hat{G}_{i + 1})}
\geq g_{\hat{c}_i, v_c}^{(r_i, \hat{G}_i)}
\geq g_{\hat{c}_k, v_c}^{(r_i, \hat{G}_i)}
%\geq g_{\hat{c}_k, v_c}^{(r_i - 1, \hat{G}_{i+1})}
\geq g_{\hat{c}_k, v_c}^{(r_{i+1}, \hat{G}_{i+1})}
\geq \cdots
\geq g_{\hat{c}_k, v_c}^{(r_k, \hat{G}_k)}
\geq g_{v_c}^{(G_{r_k + 1}^{(k)})}
= \min \left\{ \alpha_k, \beta_k \right\}
= \alpha_k\text{,}$
where the 1st and the 3rd ``$\geq$" hold because of (\ref{eqn: r-step local girth of ci}) and (\ref{eqn: r_k = min (r, dvc - k + 1)}), and the 2nd ``$\geq$" holds because of Strategy \ref{strategy: r-step M-PEGA}.
Consequently, it holds that
$g_{(\hat{c}_i, v_c)}^{(\hat{G}_{i + 1})}
= g_{\hat{c}_i, v_c}^{(r_i, \hat{G}_i)}
= g_{\hat{c}_k, v_c}^{(r_i, \hat{G}_i)}
= g_{\hat{c}_k, v_c}^{(r_k, \hat{G}_k)}
= g_{v_c}^{(G_{r_k + 1}^{(k)})}
= \alpha_k\text{.}$

As a result, on the one hand, both $\hat{c}_i$ and $\hat{c}_k$ survive the first selection criterion of Strategy \ref{strategy: r-step M-PEGA} at the beginning of  the $i$-th stage of $v_c$.
On the other hand, it holds that
\begin{equation}\label{eqn: g(ck, vc) >= beta_k > alpha_k > g(ci, vc)}
g_{(\hat{c}_k, v_c)}^{(\hat{G}_i \uplus (\hat{c}_k, v_c))}
\geq g_{(\hat{c}_k, v_c)}^{(\hat{G}_{k+1})}
\geq \max \limits_{1 \leq i \leq r_k} g_{(s_i^{(k)}, v_c)}^{(\hat{G}_k \uplus (s_i^{(k)}, v_c))}\\
\geq \max \limits_{1 \leq i \leq r_k} g_{(s_i^{(k)}, v_c)}^{(G_{i+1}^{(k)})}
\geq \beta_k
> \alpha_k
= g_{(\hat{c}_i, v_c)}^{(\hat{G}_{i + 1})}\text{,}
\end{equation}
where the 2nd ``$\geq$" holds because each CN in $\big\{\hat{c}_k, s_1^{(k)}, s_2^{(k)}, \ldots, s_{r_k}^{(k)}\big\}$ survives the first selection criterion of Strategy \ref{strategy: r-step M-PEGA} while $\hat{c}_k$ continuously survives the second selection criterion of Strategy \ref{strategy: r-step M-PEGA} at the beginning of the $k$-th stage of $v_c$.

Therefore, in the $i$-th stage of $v_c$, $\hat{c}_k$ defeats $\hat{c}_i$ with regard to the second selection criterion of Strategy \ref{strategy: r-step M-PEGA}.
Consequently, $\hat{c}_i$ has no chance to be selected for establishing the $i$-th edge of $v_c$, which results in a contradiction.
So, the assumption of $\alpha_k < \beta_k$ must be false, indicating that $\min \left\{ \alpha_k, \beta_k \right\} = \beta_k$. Till now, Proposition \ref{prop: gvc > beta} has been proved.
\end{IEEEproof}

\begin{corollary}\label{coro: gvc = beta}
For $1 \leq k \leq d_{v_c}$, if $\mathbf{S}^{(k)}$ is optimal for $g_{v_c}^{(r_k, \hat{G}_k)}$, it holds that
$g_{v_c}^{(r_k, \hat{G}_k)}
= g_{\hat{c}_k, v_c}^{(r_k, \hat{G}_k)}
= g_{v_c}^{(G_{r_k + 1}^{(k)})}
= \min \left\{ \alpha_k, \beta_k \right\}
= \beta_k\label{eqn: gvc = beta}\text{.}$
\end{corollary}

Corollary \ref{coro: gvc = beta} can be easily proved based on Proposition \ref{prop: gvc > beta}.
In addition, Proposition \ref{prop: gvc > beta} implies that the final local girth of $v_c$ is always sufficiently upper-bounded by the realtime local girths of the edges which are established in $v_c$'s later stages. Meanwhile, Corollary \ref{coro: gvc = beta} implies that currently selecting a CN, whose multi-edge local girth is optimal, is expected to have a positive effect on the final local girth of $v_c$, and this is exactly what the first selection criterion of Strategy \ref{strategy: r-step M-PEGA} does for.
To conveniently calculate the multi-edge local girths, the following corollaries are proposed.

%%%%%%%%%%%%%%%% corollary 5 %%%%%%%%%%%%%%%%%%%%
\begin{corollary}
\label{coro: easy calculation of g(vc, r)}
Assume that the current TG setting $G$ has been constructed by the $r$-edge M-PEGA, and let $\mathbf{S}_{v_c}^{(r, G)} = (s_i)_{1 \leq i \leq r}$ be a length-$r$ valid CNS with its corresponding TGS $(G_i^{(\mathbf{S})})_{1 \leq i \leq r+1}$, then it holds that
\begin{align}
g_{v_c}^{(r, G)}
&= \max \limits_{\mathbf{S}_{v_c}^{(r, G)}} \left( \min \limits_{1 \leq i \leq r} g_{(s_i, v_c)}^{(G_{i + 1}^{(\mathbf{S})})} \right)\label{eqn: gvr = max min g(ci, r)}\text{,}\\
g_{v_c}^{(r, G)}
&= \max \limits_{\mathbf{S}_{v_c}^{(r, G)}} \left( \min \limits_{1 \leq i \leq r}  \left( f_{s_i, v_c}^{(G_i^{(\mathbf{S})})} + \boldsymbol{1} \right) \right)\label{eqn: gvr = max min fcv + 1}\text{.}
\end{align}
\end{corollary}

\begin{corollary}
\label{coro: easy calculation of g(vc, 1)}
Assuming that the current TG setting $G$ has been constructed by the one-edge M-PEGA, it holds that
$g_{v_c}^{(1, G)}
 = \max \limits_{0 \leq i < m, (c_i, v_c) \notin E} \: g_{(c_i, v_c)}^{(G \uplus (c_i, v_c))}
 = \max \limits_{0 \leq i < m} \left( f_{c_i, v_c}^{(G)} + \boldsymbol{1} \right)\text{.}$
\end{corollary}

Corollary \ref{coro: easy calculation of g(vc, r)} can be easily proved based on (\ref{eqn: r-step local girth of vc}) and Corollary \ref{coro: gvc = beta}, while Corollary \ref{coro: easy calculation of g(vc, 1)} is directly deduced from Corollary \ref{coro: easy calculation of g(vc, r)}. Consequently, the one-edge M-PEGA is equivalent to the M-PEGA when the metrics are defined as the same. Furthermore, this paper employs a DFS like algorithm, presented in Algorithm \ref{algo: DFS calc in MM-PEGA},  to calculate the $r$-edge local girths in the way of (\ref{eqn: gvr = max min fcv + 1}).

\begin{algorithm}[h]
\caption{DFS Calculation of the $r$-Edge Local Girths}
\label{algo: DFS calc in MM-PEGA}
\begin{algorithmic}[1]
\REQUIRE $t, G_t = (V_c \cup V_v, E_t), g_t, u_t$.
\STATE $\lambda_t = -\boldsymbol{\infty}$.
\FOR {$i = 0; ~i < u_t; ~i +\!+$}\label{code: algo DFS @ for i}
    \IF {$(c_i, v_c) \notin E_t$}
        \STATE $G_{t+1} = G_t \uplus (c_i, v_c)$.\label{code: algo DFS @ next TG}
        \STATE $g_{(c_i, v_c)}^{(G_{t+1})} = f_{c_i, v_c}^{(G_t)} + \boldsymbol{1}$.\label{code: algo DFS @ local girth of edge}
        \STATE $g_{t+1} = \min \Big\{ g_t, g_{(c_i, v_c)}^{(G_{t+1})} \Big\}$.
        \IF {$g_{t+1} \geq g_{v_c}^{(r, G)}$}\label{code: algo DFS @ cut down useless search}
            \IF {$t == r$}
                \STATE $g_{v_c}^{(r, G)} = g_{c_i, v_c}^{(r, G)} = \lambda_t = g_{t+1}$.
            \ELSE
                \STATE $u_{t+1} = i$.
                %\STATE $G_{t+1} = \big(V, E_t \cup (c_i, v_c)\big)$.
                \STATE Calculate $F_{V_c, v_c}^{(G_{t+1})}$.\label{code: algo DFS @ calc metrics}
                \STATE Enter the ($t+1$)-th layer with input parameters $t+1, G_{t+1}, g_{t+1}, u_{t+1}$ respectively, and return here after the calculation in the ($t+1$)-th layer is finished.
                \STATE $\lambda_t = \max \Big\{ \lambda_t, \lambda_{t+1} \Big\}$.
                \STATE $g_{c_i, v_c}^{(r, G)} = \max \Big\{ g_{c_i, v_c}^{(r, G)}, \lambda_{t+1} \Big\}$.
            \ENDIF
        \ENDIF
    \ENDIF
\ENDFOR
\end{algorithmic}
\end{algorithm}

\begin{remark}
Before starting Algorithm \ref{algo: DFS calc in MM-PEGA}, $g_{v_c}^{(r, G)}$ and $g_{c_i, v_c}^{(r, G)}$, $0 \leq i < m$ need to be set as $-\boldsymbol{\infty}$.
\end{remark}

\begin{remark}
In Algorithm \ref{algo: DFS calc in MM-PEGA}, $t, G_t, g_t, u_t$ are local variables of the $t$-th (starting from 1) layer, where $t = 1, G_1 = G, g_1 = \boldsymbol{\infty}, u_1 = m$ in the first layer. Besides, other variables are global except the $i$ in line \ref{code: algo DFS @ for i}.
\end{remark}

\begin{remark}
$u_t$ works as an index-upper-bound of the enumerated CNs of the $t$-th layer, making the indices of CNs in each enumerated CNS strictly decreasing. In such case, redundant enumerations of CNSs have been avoided effectively.
\end{remark}
%\vspace{-1mm}

\begin{remark}
Line \ref{code: algo DFS @ cut down useless search} of Algorithm \ref{algo: DFS calc in MM-PEGA} cuts down useless search for $g_{v_c}^{(r, G)}$, making the DFS process much more time-efficient according to the simulation results.
\end{remark}

\begin{remark}
\label{rmk: the complexity of BFS-based MM-PEGA}
The metric-calculations between $V_c$ and $v_c$ under the realtime TG settings (i.e., those in line \ref{code: algo PEG @ calc metrics} of Algorithm \ref{algo: PEG} and in line \ref{code: algo DFS @ calc metrics} of Algorithm \ref{algo: DFS calc in MM-PEGA}) are the most time-consuming parts.
On the one hand, before each time of entering the $t$-th, $1 \leq t \leq r$ layer, the metric-calculation in the ($t-1$)-th layer\footnote{The metric-calculation in the $0$-th layer indicates that in the line \ref{code: algo PEG @ calc metrics} of Algorithm \ref{algo: PEG}.} will be implemented once.
On the other hand, the DFS process of Algorithm \ref{algo: DFS calc in MM-PEGA} will reach  the $t$-th, $1 \leq t \leq r$ layer at most $m^{t-1} / (t-1)! \approx m^{t-1}$ (since generally $r$ is small and $r \ll m$) times.
In such case, to apply Algorithm \ref{algo: DFS calc in MM-PEGA} in the $r$-edge M-PEGA once, the metric-calculation will be totally implemented at most $\sum_{t = 1}^{r} m^{t-1} = (m^r - 1) / (m - 1) \approx m^{r-1}$ (since generally $m \gg 1$) times.
In addition, the BFS like method used in the PEG algorithm \cite{Hu05} is employed to implement the metric-calculation.
Since Algorithm \ref{algo: DFS calc in MM-PEGA} will be applied in the $r$-edge M-PEGA $|E|$ times, thus the total computational complexity of the $r$-edge M-PEGA is $O\big(m^r|E| + m^{r-1}|E|^2\big)$.
\end{remark}

\section{Multi-Edge Metric-Constrained QC-PEG Algorithm}\label{sect: MM-QC-PEGA}

To design the QC-LDPC codes with any predefined valid design parameters (refer to Remark \ref{rmk: valid parameters}), as well as to detect and even to avoid generating the undetectable cycles in the QC-LDPC codes designed by the QC-PEGA \cite{	Li04} and the CP-PEGA \cite{	Lin08}, the MM-QC-PEGA is proposed in this section, which is implemented under the framework of Algorithm \ref{algo: QC-PEG} but adopts a different selection strategy.
Since the selection strategy of the MM-QC-PEGA is closely related to that of the MM-PEGA, as well as to simplify the illustration of the MM-QC-PEGA, the illustration of the MM-PEGA in last section has been employed here.
In general, because of (\ref{eqn: single edge corresponds to cyclic edges}), each concept related to a single edge $(c_i, v_j)$ in the MM-PEGA changes to a similar concept related to the edge-set $\big\{(c_i, v_c)_N\big\}$ in the MM-QC-PEGA, while the other concepts remain the same.

To be specific, the definition for a length-$r$ valid CNS $\mathbf{S} = (s_i)_{1 \leq i \leq r}$ remains the same, while its corresponding TGS $(G_i)_{1 \leq i \leq r + 1}$ changes, where $G_1 = G$ too, but $G_{i + 1} = G_i \uplus \big\{(s_i, v_c)_N\big\}$ for $1 \leq i \leq r$.
In such case, Definition \ref{def: definition of multi-edge local girths} changes a little, where $G \uplus (c_i, v_c)$ in (\ref{eqn: r-step local girth of ci}) changes to $G \uplus \big\{(c_i, v_c)_N\big\}$.
Then,  (\ref{eqn: gvr = max gcr}) still holds for the MM-QC-PEGA.
In addition, the edge-trials of the MM-QC-PEGA can be similarly defined as that in Definition \ref{def: definition of MM-PEGA, edge-trials}.
Finally, Strategy \ref{strategy: r-step M-PEGA} with two modifications on its second selection criterion, where $G \uplus (c_i, v_c)$ changes to $G \uplus \big\{(c_i, v_c)_N\big\}$ and $G \uplus (x, v_c)$ changes to $G \uplus \big\{(x, v_c)_N\big\}$, has been adopted by the MM-QC-PEGA.

To discuss the properties of the multi-edge local girths related to the QC-LDPC code, Proposition \ref{prop: gvc > beta} has been employed too, which can be proved in almost the same way with only two small modifications in (\ref{eqn: g(ck, vc) >= beta_k > alpha_k > g(ci, vc)}), where $\hat{G}_i \uplus (\hat{c}_k, v_c)$ changes to $\hat{G}_i \uplus \big\{(\hat{c}_k, v_c)_N\big\}$ and $\hat{G}_k \uplus (s_i^{(k)}, v_c)$ changes to $\hat{G}_k \uplus \big\{(s_i^{(k)}, v_c)_N\big\}$.
In addition, Corollary \ref{coro: gvc = beta} as well as (\ref{eqn: gvr = max min g(ci, r)}) in Corollary \ref{coro: easy calculation of g(vc, r)} still hold for the MM-QC-PEGA, but (\ref{eqn: gvr = max min fcv + 1}) may not hold any more because of the modification on the definition of the TGS.
To explain this, the following illustration is given.

Assume that $G = (V_c \cup V_v, E)$ is the TG of an arbitrary QC-LDPC code, and that $(c_i, v_j)$ is an arbitrary edge satisfying $c_i \in V_c$, $v_j \in V_v$, and $\big\{(c_i, v_j)_N\big\} \cap E = \emptyset$. In addition, let $\tilde{G} = G \uplus \big\{(c_i, v_j)_N\big\}$ and $\hat{G} = G \uplus \big( \big\{(c_i, v_j)_N\big\} \setminus \big\{(c_i, v_j)\big\} \big)$. Then, it holds that
\begin{equation}
\label{eqn: local girth of edge doesn't equal metric + 1}
g_{(c_i, v_j)}^{(\tilde{G})} = f_{c_i, v_j}^{(\hat{G})} + \boldsymbol{1} \leq f_{c_i, v_j}^{(G)} + \boldsymbol{1}\text{.}
\end{equation}
When $N = 1$,  $f_{c_i, v_j}^{(\hat{G})} = f_{c_i, v_j}^{(G)}$ holds, implying that in such case the $r$-edge M-PEGA is equivalent to the $r$-edge M-QC-PEGA.
However, when $N > 1$, $f_{c_i, v_j}^{(\hat{G})} < f_{c_i, v_j}^{(G)}$ may sometimes hold.
For example, let $G$ be the TG of Fig. \ref{fig: undetectableCycles}(a), further assume that $c_i = c_2$ and $v_j = v_0$, then $f_{c_i, v_j}^{(\hat{G})} = 3 < f_{c_i, v_j}^{(G)} = \infty$.
Consequently, on the one hand, (\ref{eqn: local girth of edge doesn't equal metric + 1}) explains the essential reasons that why the cycles, which contain two or more newly established edges, cannot be detected in the QC-LDPC codes designed by the QC-PEGA \cite{	Li04} and the CP-PEGA \cite{	Lin08}, but can be detected in the codes designed by the MM-QC-PEGA.
On the other hand, Algorithm \ref{algo: DFS calc in MM-PEGA} needs to be modified when being applied in the MM-QC-PEGA to calculate the multi-edge local girths.
Exactly, in Algorithm \ref{algo: DFS calc in MM-PEGA}, line \ref{code: algo DFS @ next TG} changes to $G_{t+1} = G_t \uplus \big\{(c_i, v_c)_N\big\}$, and line \ref{code: algo DFS @ local girth of edge} changes to
\begin{equation}
\label{eqn: calculation of local girth of edge from metric in QC}
g_{(c_i, v_c)}^{(G_{t+1})} = f_{c_i, v_c}^{(\bar{G})} + \boldsymbol{1}\text{,}
\end{equation}
where $\bar{G} = G_t \uplus \big( \big\{(c_i, v_c)_N\big\} \setminus \big\{(c_i, v_c)\big\} \big)$.

To calculate $f_{c_i, v_c}^{(\bar{G})}$ in (\ref{eqn: calculation of local girth of edge from metric in QC}), the BFS like method used in the original PEG algorithm \cite{	Hu05} is employed.
In such case, the computational complexity for calculating $f_{c_i, v_c}^{(\bar{G})}$ once is $O\big( m + |E| \big)$, while the metric-calculation in line \ref{code: algo QC-PEG @ calc metrics} of Algorithm \ref{algo: QC-PEG} as well as that in line \ref{code: algo DFS @ calc metrics} of Algorithm \ref{algo: DFS calc in MM-PEGA} can be omitted.
Since the calculation for $f_{c_i, v_c}^{(\bar{G})}$ will be implemented approximately $m^r$ times at each stage of the $r$-edge M-QC-PEGA according to Remark \ref{rmk: the complexity of BFS-based MM-PEGA}, thus the total computational complexity of the $r$-edge M-QC-PEGA is $O\big( (m^{r+1}|E| + m^r|E|^2) / N \big)$.

\section{Simulation Results}\label{sect: simulation results}

In this paper, a code is said to be regular if its VN-degrees are identical. Besides, the selection strategies of the aforementioned LDPC code design algorithms have tried to make the CN-degrees as identical as possible too.
The bit error rate (BER) estimations of all simulated codes are performed using the Monte-Carlo simulation, assuming binary phase-shift keying (BPSK) transmission over the AWGN channel and the standard sum-product algorithm (SPA) iterative decoding with 100 decoding iterations at the receiver. In addtion, at least 100 frame errors are captured at each simulated signal-to-noise ratio (SNR) point.

\subsection{Irregular LDPC Codes}

\begin{example}\label{eg: con time, ACE spectra, and BER of irreg deg15 1008 504}
This example considers the design of the irregular $(1008, 504)$ non-QC-LDPC codes using the MM-PEGA. The popular density evolution (DE) \cite{	Richardson01a, Richardson01b} optimized VN-degree distribution\footnote{Assume that the VN-degree distribution of an LDPC code is $\gamma(x) = \sum_i p_ix^i$, then $p_i$ has indicated the percentage of the degree-$i$ VNs among all the VNs.} $\gamma(x) = 0.47532x^2 + 0.27953x^3 + 0.03486x^4 + 0.10889x^5 + 0.10138x^{15}$ is employed to design the codes.
For convenience, the code (or code set) designed by the $r$-edge M-PEGA with the metric (\ref{eqn: metric as distance}) is denoted as $A_r$, and that designed by the $r$-edge M-PEGA with the metric (\ref{eqn: metric as dist and ACE pair}) is denoted as $B_r$, where $r = 1, 2, 3, 4$ for both $A_r$ and $B_r$.

%\begin{table}[!t]
%\renewcommand{\arraystretch}{1.0}
%\caption{Average Construction Time of a Single Code of Different Code Sets in Example \ref{eg: con time, ACE spectra, and BER of irreg deg15 1008 504}}
%\label{table: irreg deg15 1008 504 @ time}
%\centering
%\begin{tabular}{p{0.9cm}p{0.9cm}p{0.9cm}||p{0.9cm}p{0.9cm}p{0.9cm}}
%\hline
%\bfseries Code & \bfseries BFS & \bfseries SBM & \bfseries Code & \bfseries BFS & \bfseries SBM \\
%\hline
%$A_1$   & $0.4$s    & $2.8$s    & $B_1$     & $0.4$s    & $2.7$s  \\
%%\hline
%$A_2$   & $79.2$s   & $9.7$s    & $B_2$     & $78.1$s   & $7.8$s  \\
%%\hline
%$A_3$   & $266.9$s  & $21.2$s   & $B_3$     & $314.8$s  & $22.2$s  \\
%%\hline
%$A_4$   & $452.1$s  & $28.1$s   & $B_4$     & $779.7$s  & $45.0$s  \\
%\hline
%\end{tabular}
%\end{table}

\begin{table}[!t]
\renewcommand{\arraystretch}{1.0}
\caption{ACE Spectra of Different Code Sets in Example \ref{eg: con time, ACE spectra, and BER of irreg deg15 1008 504}}
\label{table: irreg deg15 1008 504 @ average}
\centering
\begin{tabular}{clcc}
\hline
\multicolumn{1}{c}{\bfseries Code} & \multicolumn{1}{c}{\bfseries Average} &
\multicolumn{1}{c}{\bfseries Maximum} & \multicolumn{1}{c}{\bfseries Freq} \\
\hline
$A_1$   & $(\infty, \infty, 13.00, 6.08, 3.03)$     & $(\infty, \infty, 13, 13, 4)$   & $0.0038$ \\
$A_2$   & $(\infty, \infty, 13.00, 8.20, 3.03)$     & $(\infty, \infty, 13, 13, 4)$   & $0.0130$ \\
$A_3$   & $(\infty, \infty, 13.00, 12.53, 3.03)$    & $(\infty, \infty, 13, 13, 4)$   & $0.0326$ \\
$A_4$   & $(\infty, \infty, 13.00, 13.00, 3.03)$    & $(\infty, \infty, 13, 13, 4)$   & $0.0295$ \\
$B_1$   & $(\infty, \infty, 18.21, 8.85, 3.82)$     & $(\infty, \infty, 26, 10, 4)$   & $0.0012$ \\
$B_2$   & $(\infty, \infty, 18.53, 9.61, 3.88)$     & $(\infty, \infty, 26, 13, 5)$   & $0.0006$ \\
$B_3$   & $(\infty, \infty, 19.12, 12.69, 4.66)$    & $(\infty, \infty, 26, 13, 5)$   & $0.0313$ \\
$B_4$   & $(\infty, \infty, 20.44, 13.00, 5.08)$    & $(\infty, \infty, 26, 13, 6)$   & $0.0229$ \\
\hline
\end{tabular}
\end{table}

\begin{table}[!t]
\renewcommand{\arraystretch}{1.0}
\caption{ACE Spectra of the Selected Codes in Example \ref{eg: con time, ACE spectra, and BER of irreg deg15 1008 504} and the Frequencies Associated With the Former ACE Spectra}
\label{table: irreg deg15 1008 504 @ selected}
\centering
\begin{tabular}{ccc}
\hline
\multicolumn{1}{c}{\bfseries Code} & \multicolumn{1}{c}{\bfseries Spectrum} & \multicolumn{1}{c}{\bfseries Freq} \\
\hline
$A_1$   & $(\infty, \infty, 13, 13, 3)$    & $0.1229$ \\
$A_2$   & $(\infty, \infty, 13, 13, 4)$    & $0.0130$ \\
$A_3$   & $(\infty, \infty, 13, 13, 4)$    & $0.0326$ \\
$A_4$   & $(\infty, \infty, 13, 13, 4)$    & $0.0295$ \\
$B_1$   & $(\infty, \infty, 19, 10, 4)$    & $0.0259$ \\
$B_2$   & $(\infty, \infty, 19, 13, 4)$    & $0.0128$ \\
$B_3$   & $(\infty, \infty, 26, 13, 5)$    & $0.0313$ \\
$B_4$   & $(\infty, \infty, 26, 13, 6)$    & $0.0229$ \\
\hline
\end{tabular}
\vspace{-3mm}
\end{table}

%Firstly, to estimate the efficiency of the SBM, 10 codes of each code set  are randomly constructed using a desktop computer with an Intel Core i7-2600 3.4-GHz CPU and 4GB-memory.
%The average construction time for a single code from the different code sets are presented in Table \ref{table: irreg deg15 1008 504 @ time}, where it is verified that the set-based $r$-edge M-PEGA is much more time-efficient than the BFS-based $r$-edge M-PEGA when the edge-trials $r > 1$.

More than 1000 codes of each code set are randomly constructed. The average ACE spectrum, the maximum ACE spectrum, and the frequency\footnote{The frequency associated with an ACE spectrum indicates the frequency that the assumed ACE spectrum occurs among all the spectra of the codes from a same code set.} associated with the maximum ACE spectrum of each code set  are presented in Table \ref{table: irreg deg15 1008 504 @ average}.
In Table \ref{table: irreg deg15 1008 504 @ average}, for the code sets designed with the metric (\ref{eqn: metric as distance}) (i.e., $A_1$--$A_4$), it can be seen that the average ACE spectrum increases as the edge-trials increases, but different code sets achieve the same maximum ACE spectrum. In general, the code set  with a larger edge-trials can achieve the maximum ACE spectrum with a higher frequency.
At the same time, for the code sets designed with the metric (\ref{eqn: metric as dist and ACE pair}) (i.e., $B_1$--$B_4$), it can be seen that both the average and the maximum ACE spectra increase as the edge-trials increases.
In addition, it's consistent with the simulation results in \cite{	Vukobratovic08} that the ACE spectra of the codes designed with the metric (\ref{eqn: metric as dist and ACE pair}) greatly surpass that of the codes designed with the metric (\ref{eqn: metric as distance}).
Furthermore, compared to the ACE spectra of the LDPC codes constructed with the similar design parameters in the literature \cite{	Asvadi12, Vukobratovic07, Vukobratovic08, Vukobratovic09}, the maximum ACE spectra of $B_1$--$B_4$ reported in Table \ref{table: irreg deg15 1008 504 @ average} have been the best till now.

\begin{figure}[!t]
\centering
\includegraphics[scale = 0.5]{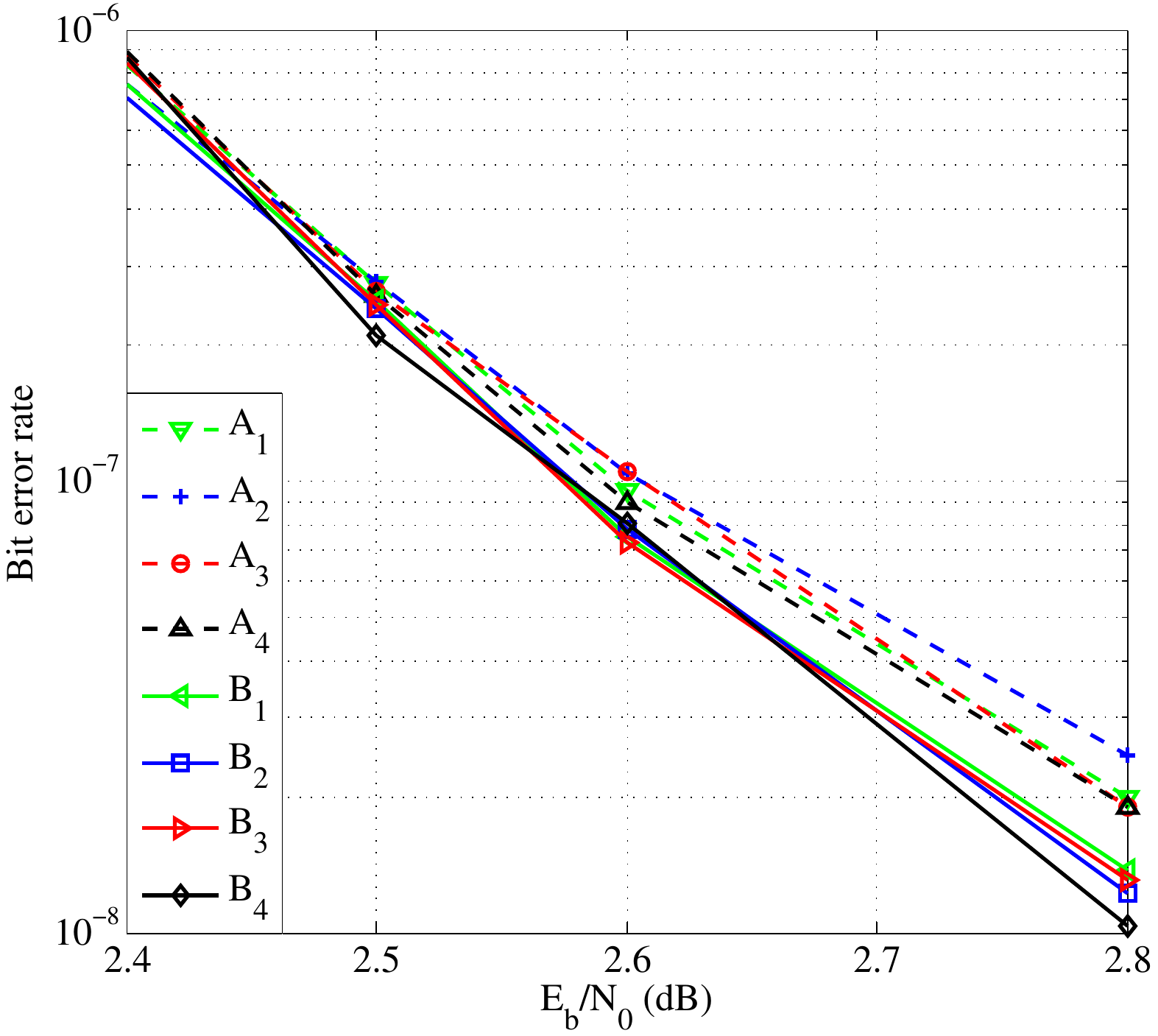}
\vspace{-3mm}
\caption{BER performances of the irregular $(1008, 504)$ LDPC codes designed in Example \ref{eg: con time, ACE spectra, and BER of irreg deg15 1008 504}.}
\label{fig: BER irreg deg15 1008 504}
\vspace{-3mm}
\end{figure}

Two codes are randomly selected from each code set, keeping the ACE spectrum of each selected code maximum with an associated frequency larger than $0.01$\footnote{This threshold is chosen with two considerations: Firstly, each code set  offers at least 10 candidate codes; Secondly, the ACE spectrum should be large enough so that the chance to get a code with the same ACE spectrum by reconstructing codes of the same code set  will not be too small.}.
The ACE spectra of these selected codes as well as the frequencies associated with the former ACE spectra are presented in Table \ref{table: irreg deg15 1008 504 @ selected}.
In addition, the BER estimations of these codes are performed.
Then, for the two codes from a same code set, the one which has achieved a lower BER at $E_b/N_0 = 2.8$dB  is remained.
Furthermore, the error performances of the remained codes are presented in Fig. \ref{fig: BER irreg deg15 1008 504}.
From Fig. \ref{fig: BER irreg deg15 1008 504}, it can be seen that each code designed with the metric (\ref{eqn: metric as dist and ACE pair}) outperforms all that codes designed with the metric (\ref{eqn: metric as distance}) in the high SNR region between $2.6$dB and $2.8$dB.
In addition, among the four codes designed with the metric (\ref{eqn: metric as dist and ACE pair}) (i.e., $B_1$--$B_4$), $B_4$ with the best ACE spectrum achieves the lowest BER, and $B_1$ with the poorest ACE spectrum holds the highest BER.
In such case, the simulation results in Fig. \ref{fig: BER irreg deg15 1008 504} are consistent with that increasing the ACE spectrum is expected to have a positive effect on the error performance \cite{Tian04, Xiao04, Vukobratovic07, Vukobratovic08, Vukobratovic09}.

\end{example}

\begin{example}\label{eg: irreg deg15 1008 504 QC}
This example considers the design of the irregular $(1008, 504)$ QC-LDPC codes with the one-dimension circulant-size $N = 36$ and the VN-degree distribution $\gamma(x) = 0.46429x^2 + 0.28571x^3 + 0.03571x^4 + 0.10714x^5 + 0.10714x^{15}$ using the MM-QC-PEGA. Besides, $N$ cannot be too large, or the VN-degree distribution will change heavily from the DE-optimized one used in Example \ref{eg: con time, ACE spectra, and BER of irreg deg15 1008 504}.
For convenience, denote $C_{\alpha, \beta}, \alpha = 1, 2, \beta = 0, 1, 2$ as the code (or code set) designed by the QC-PEGA \cite{Li04} with the metric ($\alpha$) when $\beta = 0$, and as the code (or code set) designed by the $\beta$-edge M-QC-PEGA with the metric ($\alpha$) when $\beta = 1$ or $2$.
However, the CP-PEGA \cite{	Lin08}, which requires each nonzero circulant of the check matrix to be a CPM, does not suit for this example.

\begin{table}[!t]
\renewcommand{\arraystretch}{1.0}
\caption{ACE Spectra of Different Code Sets in Example \ref{eg: irreg deg15 1008 504 QC}}
\label{table: irreg deg15 1008 504 QC}
\centering
\begin{tabular}{cllc}
\hline
\multicolumn{1}{c}{\bfseries Code} & \multicolumn{1}{c}{\bfseries Average} &
\multicolumn{1}{c}{\bfseries Maximum} & \multicolumn{1}{c}{\bfseries Freq} \\
\hline
$C_{1, 0}$   & $(\infty, \;\text{-}\textsuperscript{a} , 8.30,	4.94,	2.80)$         & $(\infty, \infty, 15, 2, 4)$   & $0.001$ \\
$C_{1, 1}$   & $(\infty, \infty, 13.39,	12.63,	4.01)$     & $(\infty, \infty, 14, 14, 5)$   & $0.015$ \\
$C_{1, 2}$   & $(\infty, \infty, 13.45,	13.03,	4.06)$     & $(\infty, \infty, 14, 14, 5)$   & $0.014$ \\
$C_{2, 0}$   & $(\infty, \;\text{-}\textsuperscript{b} , 9.19,	3.98,	2.74)$         & $(\infty, \infty, 26, 14, 6)$   & $0.001$ \\
$C_{2, 1}$   & $(\infty, \infty, 24.32,	12.71,	4.89)$     & $(\infty, \infty, 26, 14, 6)$   & $0.021$ \\
$C_{2, 2}$   & $(\infty, \infty, 24.71,	13.20,	5.08)$     & $(\infty, \infty, 26, 14, 6)$   & $0.048$ \\
\hline
\multicolumn{4}{r}{\textsuperscript{a}46\% codes of this code set  are free of 4-cycles.}\\
\multicolumn{4}{r}{\textsuperscript{b}13\% codes of this code set  are free of 4-cycles.}\\
\end{tabular}
\vspace{-3mm}
\end{table}

Exact 1000 QC-LDPC codes of each code set  are randomly constructed. In addition, the average ACE spectrum, the maximum ACE spectrum, and the frequency associated with the maximum ACE spectrum of each code set  are presented in Table \ref{table: irreg deg15 1008 504 QC}.
The average ACE spectra in Table \ref{table: irreg deg15 1008 504 QC} show that the QC-PEGA \cite{Li04} sometimes results in 4-cycles while the MM-QC-PEGA avoid generating 4-cycles effectively.
In addition, the MM-QC-PEGA overwhelmingly outperforms the QC-PEGA \cite{Li04} in terms of the average ACE spectra, indicating that the MM-QC-PEGA is much more effective than the QC-PEGA \cite{Li04} for designing the QC-LDPC code with larger girth as well as better ACE spectrum.
However, one code in the code set  $C_{1, 0}$ achieves the largest ACE spectrum $(\infty, \infty, 15, 2, 4)$ among the code sets $C_{1, \beta}, \beta = 0, 1, 2$. But this code may hold a high error floor since there apparently contains some small ETSs, such as the $(4, 2)$ ETS and the $(5, 4)$ ETS, each of which consists of a single cycle.
Instead, the second largest ACE spectrum of the code set  $C_{1, 0}$, i.e., $(\infty, \infty, 14, 14, 5)$, which equals the maximum ACE spectra of $C_{1, \beta}, \beta = 1, 2$ but occurs in a much smaller frequency $0.001$, should be considered as a better one.
Furthermore, for the codes designed by the MM-QC-PEGA, their ACE spectra under the metric (\ref{eqn: metric as dist and ACE pair}) greatly surpass that under the metric (\ref{eqn: metric as distance}), which is also consistent with the simulation results in Example \ref{eg: con time, ACE spectra, and BER of irreg deg15 1008 504} and that in \cite{	Vukobratovic08}.
Meanwhile, under the same metric, the average ACE spectrum achieved by the two-edge M-QC-PEGA is slightly better than that achieved by the one-edge M-PEGA.
When compare Table \ref{table: irreg deg15 1008 504 QC} to Table \ref{table: irreg deg15 1008 504 @ average}, the average and the maximum ACE spectra achieved by the MM-QC-PEGA in  Table \ref{table: irreg deg15 1008 504 QC} are better than that achieved by the MM-PEGA in  Table \ref{table: irreg deg15 1008 504 @ average}.\footnote{As the VN-degree distributions used in Example \ref{eg: con time, ACE spectra, and BER of irreg deg15 1008 504} and Example \ref{eg: irreg deg15 1008 504 QC} only differ a little from each other, the comparison between Table \ref{table: irreg deg15 1008 504 @ average} (Example \ref{eg: con time, ACE spectra, and BER of irreg deg15 1008 504}) and Table \ref{table: irreg deg15 1008 504 QC} (Example \ref{eg: irreg deg15 1008 504 QC}) is quite fair. Also, to verify this, under the VN-degree distribution used in Example \ref{eg: irreg deg15 1008 504 QC}, another 1000 codes have been randomly constructed using each of the MM-PEGA with edge-trials 1 and 2 and with metrics (\ref{eqn: metric as distance}) and (\ref{eqn: metric as dist and ACE pair}). Then, the statistical results of these codes show that both their average and maximum ACE spectra are inferior to the corresponding ones of Table \ref{table: irreg deg15 1008 504 @ average}.}
In such case, with regard to designing the irregular LDPC codes with better ACE spectra, the QC-LDPC codes, which have proper circulant-sizes and are designed by the MM-QC-PEGA, may be more preferable than the non-QC-LDPC codes designed by the MM-PEGA.

\begin{figure}[!t]
\centering
\includegraphics[scale = 0.5]{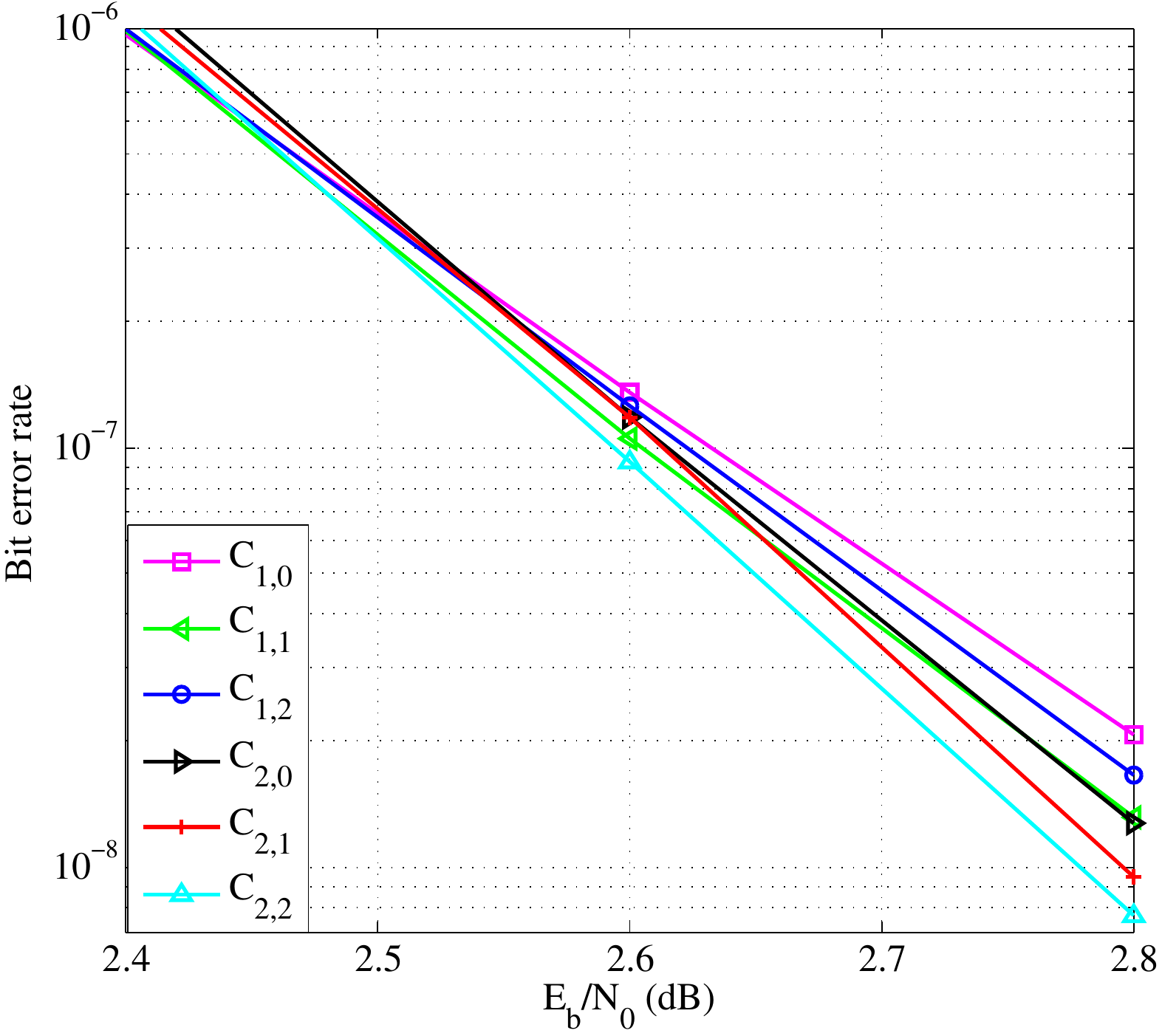}
\vspace{-3mm}
\caption{BER performances of the irregular $(1008, 504)$ QC-LDPC codes with $N = 36$ designed in Example \ref{eg: irreg deg15 1008 504 QC}.}
\label{fig: BER irreg deg15 1008 504 QC}
\vspace{-3mm}
\end{figure}

To perform the BER estimations, one of the codes whose ACE spectra are the maximum among the first 100 constructed codes of each code set  is randomly selected.\footnote{This is because the error estimations in this example have been performed after 100 codes of each code set  are constructed. However, to make the aforementioned investigation of the ACE spectra more accurate, another 900 codes of each code set  have been constructed. Furthermore, in the rest of this paper, the codes for performing error estimations are always selected from 100 candidate codes.}
The ACE spectrum of $C_{1, 0}$ is $(\infty, \infty, 14, 14, 3)$, and that of $C_{1, 2}$ is $(\infty, \infty, 14, 14, 4)$, and that of the other codes remain the same as the maximum ACE spectra of their own code sets in Table \ref{table: irreg deg15 1008 504 QC}.
The BER performances of these selected codes are presented in Fig. \ref{fig: BER irreg deg15 1008 504 QC}, where the BER performances of these codes match their  ACE spectra well, i.e. the code with a better ACE spectrum achieves a lower BER.
Furthermore, both $C_{2, 1}$ and $C_{2, 2}$ slightly outperform the best code in Fig. \ref{fig: BER irreg deg15 1008 504} (i.e., $B_4$) with regard to the BER at $E_b/N_0 = 2.8$dB.
On the one hand, this is interesting since $C_{2, 1}$ and $C_{2, 2}$ are QC while $B_4$ is non-QC.
On the other hand, it is reasonable since $C_{2, 1}$ and $C_{2, 2}$ hold a better ACE spectrum compared to that of $B_4$.
In summary, the simulation results in this example are also consistent with that better ACE spectra generally result in lower error rates \cite{Tian04, Xiao04, Vukobratovic07, Vukobratovic08, Vukobratovic09}.

\end{example}

\subsection{Regular LDPC Codes}

\begin{table}[!t]
\renewcommand{\arraystretch}{0.82}
\caption{VNLGDs of Different QC-LDPC Codes in Example \ref{eg: reg deg3 1024 512 QC}}
\label{table: VNLGDs of the QC-LDPC codes}
\centering

\subtable[QC-PEGA\cite{	Li04}]{
\begin{tabular}{llll}
\hline
\multicolumn{1}{c}{$N$} &   \multicolumn{1}{c}{\bfseries Average} & \multicolumn{1}{c}{\bfseries Minimum} & \multicolumn{1}{c}{\bfseries Freq} \\
\hline
$2^0$   &   $0.1058x^4 + 0.8942x^6$     &   $0.0293x^8 + 0.9707x^{10}$  &   $0.01$   \\
$2^1$   &   $0.0078x^4 + 0.1628x^8 + 0.8294x^{10}$              &   $0.1426x^8 + 0.8574x^{10}$  &   $0.01$   \\
$2^2$   &   $0.0080x^4 + 0.0004x^6  + 0.1619x^8 + 0.8297x^{10}$ &   $0.1250x^8 + 0.8750x^{10}$  &   $0.01$   \\
$2^3$   &   $0.0092x^4 + 0.0018x^6  + 0.1727x^8 + 0.8163x^{10}$ &   $0.0625x^8 + 0.9375x^{10}$  &   $0.01$   \\
$2^4$   &   $0.0095x^4 + 0.0053x^6  + 0.1914x^8 + 0.7938x^{10}$ &   $0.0312x^8 + 0.9688x^{10}$  &   $0.04$   \\
$2^5$   &   $0.0160x^4 + 0.0096x^6  + 0.2078x^8 + 0.7666x^{10}$ &   $1.0000x^{10}$  &   $0.02$   \\
$2^6$   &   $0.0325x^4 + 0.0619x^6  + 0.2694x^8 + 0.6362x^{10}$ &   $1.0000x^{10}$  &   $0.01$   \\
$2^7$   &   $0.1500x^4 + 0.1525x^6  + 0.3500x^8 + 0.3475x^{10}$ &   $1.0000x^{10}$  &   $0.03$   \\
$2^8$   &   $0.3825x^4 + 0.1625x^8  + 0.4550x^8$                &   $1.0000x^{8}$   &   $0.22$   \\
$2^9$   &   $0.4500x^4 + 0.5500x^6$                             &   $1.0000x^{6}$   &   $0.38$   \\\hline
\end{tabular}
\label{subtable: VNLGDs of QC-PEGA}
}

\subtable[CP-PEGA\cite{	Lin08}]{
\begin{tabular}{llll}
\hline
\multicolumn{1}{c}{$N$} &   \multicolumn{1}{c}{\bfseries Average} & \multicolumn{1}{c}{\bfseries Minimum} & \multicolumn{1}{c}{\bfseries Freq} \\
\hline
$2^0$ & $0.1058x^8 + 0.8942x^{10}$  &   $0.0293x^8 + 0.9707x^{10}$  &   $0.01$  \\
$2^1$ & $0.1748x^8 + 0.8252x^{10}$  &   $0.0566x^8 + 0.9434x^{10}$  &   $0.01$  \\
$2^2$ & $0.1732x^8 + 0.8268x^{10}$  &   $0.0312x^8 + 0.9688x^{10}$  &   $0.01$  \\
$2^3$ & $0.1704x^8 + 0.8296x^{10}$  &   $0.0312x^8 + 0.9688x^{10}$  &   $0.01$  \\
$2^4$ & $0.2133x^8 + 0.7867x^{10}$  &   $1.0000x^{10}$              &   $0.02$  \\
$2^5$ & $0.2622x^8 + 0.7378x^{10}$  &   $1.0000x^{10}$              &   $0.09$  \\
$2^6$ & $0.4044x^8 + 0.5956x^{10}$  &   $1.0000x^{10}$              &   $0.12$  \\
$2^7$ & $0.5900x^8 + 0.4100x^{10}$  &   $1.0000x^{10}$              &   $0.06$  \\
\hline
\end{tabular}
}

\subtable[One-Edge M-QC-PEGA]{
\begin{tabular}{llll}
\hline
\multicolumn{1}{c}{$N$} &   \multicolumn{1}{c}{\bfseries Average} & \multicolumn{1}{c}{\bfseries Minimum} & \multicolumn{1}{c}{\bfseries Freq} \\
\hline
$2^0$ & $0.1058x^8 + 0.8942x^{10}$  &   $0.0293x^8 + 0.9707x^{10}$  &   $0.01$  \\
$2^1$ & $0.1303x^8 + 0.8697x^{10}$  &   $0.0195x^8 + 0.9805x^{10}$  &   $0.01$  \\
$2^2$ & $0.1196x^8 + 0.8804x^{10}$  &   $1.0000x^{10}$              &   $0.03$  \\
$2^3$ & $0.1375x^8 + 0.8625x^{10}$  &   $1.0000x^{10}$              &   $0.15$  \\
$2^4$ & $0.1620x^8 + 0.8380x^{10}$  &   $1.0000x^{10}$              &   $0.32$  \\
$2^5$ & $0.1191x^8 + 0.8809x^{10}$  &   $1.0000x^{10}$              &   $0.63$  \\
$2^6$ & $0.1731x^8 + 0.8269x^{10}$  &   $1.0000x^{10}$              &   $0.70$  \\
$2^7$ & $0.1225x^8 + 0.8775x^{10}$  &   $1.0000x^{10}$              &   $0.85$  \\
$2^8$ & $1.0000x^8$                 &   $1.0000x^8$                 &   $1.00$  \\
$2^9$ & $1.0000x^6$                 &   $1.0000x^6$                 &   $1.00$  \\
\hline
\end{tabular}
\label{subtable: VNLGDs of 1SMCPEGQCA}
}

\subtable[Two-Edge M-QC-PEGA]{
\begin{tabular}{llll}
\hline
\multicolumn{1}{c}{$N$} &   \multicolumn{1}{c}{\bfseries Average} & \multicolumn{1}{c}{\bfseries Minimum} & \multicolumn{1}{c}{\bfseries Freq} \\
\hline
$2^0$ & $0.0004x^8 + 0.9996x^{10}$  &   $1.0000x^{10}$              &   $0.97$  \\
$2^1$ & $0.0003x^8 + 0.9997x^{10}$  &   $1.0000x^{10}$              &   $0.99$  \\
$2^2$ & $0.0006x^8 + 0.9994x^{10}$  &   $1.0000x^{10}$              &   $0.99$  \\
$2^3$ & $0.0017x^8 + 0.9983x^{10}$  &   $1.0000x^{10}$              &   $0.98$  \\
$2^4$ & $0.0017x^8 + 0.9983x^{10}$  &   $1.0000x^{10}$              &   $0.99$  \\
$2^5$ & $1.0000x^{10}$              &   $1.0000x^{10}$              &   $1.00$  \\
$2^6$ & $1.0000x^{10}$              &   $1.0000x^{10}$              &   $1.00$  \\
$2^7$ & $1.0000x^{10}$              &   $1.0000x^{10}$              &   $1.00$  \\
$2^8$ & $1.0000x^8$                 &   $1.0000x^8$                 &   $1.00$  \\
$2^9$ & $1.0000x^6$                 &   $1.0000x^6$                 &   $1.00$  \\
\hline
\end{tabular}
\label{subtable: VNLGDs of 2SMCPEGQCA}
}
\vspace{-2cm}
\end{table}

\begin{example}\label{eg: reg deg3 1024 512 QC}
This example considers the design of the regular $(1024, 512)$ QC-LDPC codes with VN-degree-three. For each valid one-dimension circulant-size $N$ from $\left\{2^0, 2^1, \ldots, 2^9 \right\}$ and each design algorithm of the QC-PEGA \cite{Li04}, the CP-PEGA \cite{	Lin08}, the one-edge M-QC-PEGA, and the two-edge M-QC-PEGA, exact 100 QC-LDPC codes are randomly constructed.\footnote{When constructing regular LDPC codes, it makes no difference to use metric (\ref{eqn: metric as distance}) or metric (\ref{eqn: metric as dist and ACE pair}).}
For each code set of 100 codes with the same $N$ and the same design algorithm, their average and minimum VNLGDs are presented in Table \ref{table: VNLGDs of the QC-LDPC codes}.
At the same time, the frequency that the minimum VNLGD occurs among all the VNLGDs of the codes from a same code set  is given in Table \ref{table: VNLGDs of the QC-LDPC codes} too.
For each specific $N$  from Table \ref{subtable: VNLGDs of QC-PEGA} to Table \ref{subtable: VNLGDs of 2SMCPEGQCA}, both the average and the minimum VNLGDs keep nonincreasing while the frequencies keep nondecreasing,\footnote{$N = 2^4$ is the only one exception where the frequencies do not keep nondecreasing.} indicating that the MM-QC-PEGA, especially that with a larger edge-trials, is more effective than the QC-PEGA \cite{Li04} and the CP-PEGA \cite{	Lin08} for designing the QC-LDPC codes with better VNLGDs.
Furthermore, according to the comparison between the QC-LDPC codes with the different $N$s in the same table from Table \ref{subtable: VNLGDs of QC-PEGA} to Table \ref{subtable: VNLGDs of 2SMCPEGQCA}, the codes with moderate $N$s, such as $N = 2^5$, $2^6$, $2^7$, have larger chances to achieve the optimal VNLGDs. By the way, if $N$ is too large, such as $N = 2^8$, $2^9$, short inevitable cycles will form in the QC-LDPC code with the assumed $N$, which also coincides with the results in \cite{	Park13}.

For $N = 2^3$ and $N = 2^7$ in each table from Table \ref{subtable: VNLGDs of QC-PEGA} to Table \ref{subtable: VNLGDs of 2SMCPEGQCA}, one of the codes whose VNLGDs are the minimum is randomly selected to perform the BER estimations.
The error performances of the selected codes are presented in Fig. \ref{fig: BER reg deg3 1024_512 QC}, where the codes are labelled based on their design algorithms and their one-dimension circulant-sizes, such as 1M-QC-PEGA-$2^3$ denotes the code which is designed by the one-edge M-QC-PEGA with $N = 2^3$.
From Fig. \ref{fig: BER reg deg3 1024_512 QC}, it can be seen that the codes designed by the MM-QC-PEGA outperform that codes designed by the QC-PEGA \cite{Li04} and the CP-PEGA \cite{	Lin08} at most $0.1$dB in the high SNR region between $2.8$dB and $3.0$dB.
At the same time, the code 2M-QC-PEGA-$2^7$ slightly outperforms the regular (1024, 512) non-QC-LDPC codes designed by the MM-PEGA in Example \ref{eg: reg deg3 3 codes} at $E_b/N_0 = 3.0$dB, presenting another sample for that the QC-LDPC codes designed by the MM-QC-PEGA can sometimes outperform the non-QC-LDPC codes designed by the MM-PEGA.
%while the most QC-LDPC-code-design algorithms in \cite{	Kou01, Lan07, Li14, Myung05, Jiang09,	Jiang14, Huang10, Bocharova11, Bocharova12, Asvadi11, Asvadi12, Park13, Mitchell14, Fossorier04, Li04, Lin08} can at most achieve the BER performances as better as the PEG algorithm \cite{Hu05} (i.e., the one-edge M-PEGA).

\begin{figure}[!t]
\centering
\includegraphics[scale = 0.5]{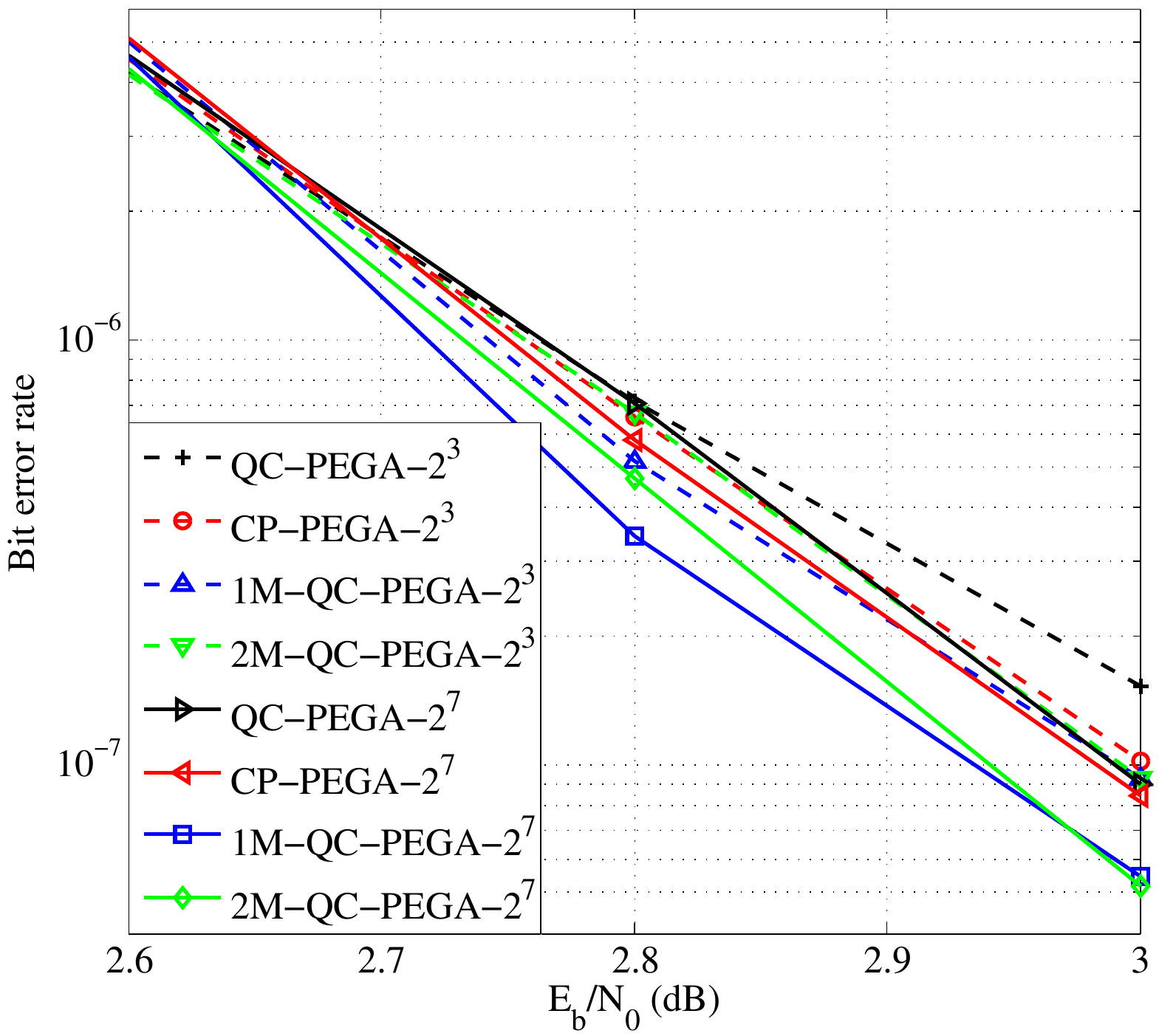}
\vspace{-3mm}
\caption{BER performances of the regular $(1024, 512)$ QC-LDPC codes with $N = 2^3$ and $N = 2^7$ designed in Example \ref{eg: reg deg3 1024 512 QC}.}
\label{fig: BER reg deg3 1024_512 QC}
\vspace{-3mm}
\end{figure}

\end{example}

\begin{figure}[!t]
\centering
\includegraphics[scale = 0.5]{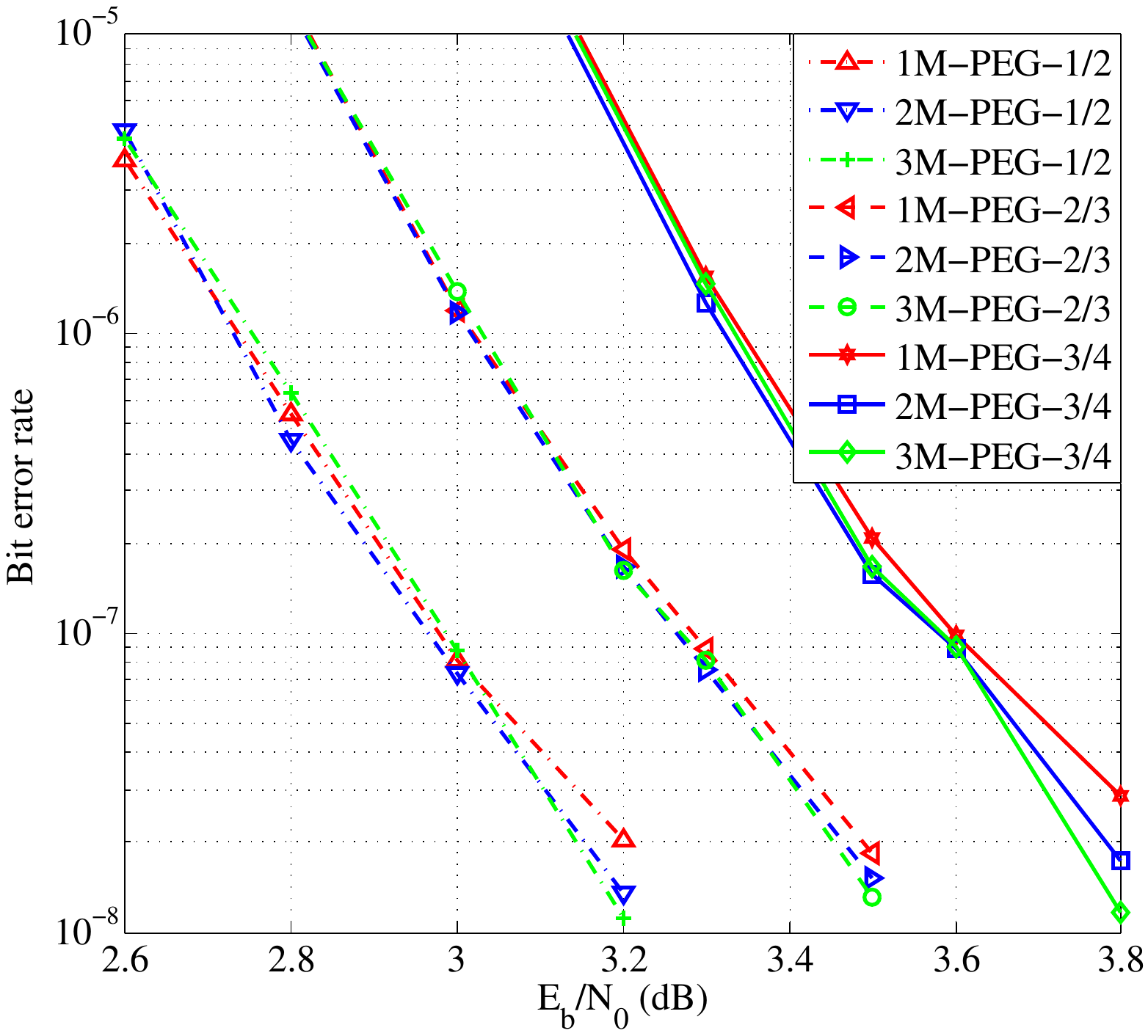}
\vspace{-3mm}
\caption{BER performances of a sequence of codes with rates $1/2$, $2/3$, and $3/4$ designed in Example \ref{eg: reg deg3 3 codes}.}
\label{fig: BER reg deg3 3 codes}
\vspace{-3mm}
\end{figure}

\begin{example}\label{eg: reg deg3 3 codes}
This example considers the design of  three different types of regular non-QC-LDPC codes with VN-degree-three using the MM-PEGA. These types of codes are the rate-$1/2$ $(1024, 512)$, the rate-$2/3$ $(1536, 1024)$, and the rate-$3/4$ $(2048, 1536)$ LDPC codes.
For each rate and each of the MM-PEGA with edge-trials 1, 2, and 3, exact 100 codes are randomly constructed.
For convenience, different codes (code sets) are denoted based on their design algorithms and rates, such as 1M-PEGA-$1/2$ denotes the code (code set) which is designed by the one-edge M-PEGA and has the rate $1/2$.
Furthermore, one of the codes whose VNLGDs are the minimum is randomly selected from each code set  to perform the BER estimations,\footnote{The codes 1M-PEGA-$1/2$ and 2M-PEGA-$1/2$ are selected from the code sets of Example \ref{eg: reg deg3 1024 512 QC}, which are designed by the MM-QC-PEGA and have one-dimension circulant-size $1$.} where the VNLGD of 1M-PEGA-$1/2$ is $0.0293x^8 + 0.9707x^{10}$, and that of the other rate-$1/2$ codes are $x^{10}$, and that of the left codes are $x^8$.
At the same time, the error performances of the selected codes are presented in Fig. \ref{fig: BER reg deg3 3 codes}.
Furthermore, according to the comparisons among the three codes with the same rates, it's verified that increasing the edge-trials of the MM-PEGA is expected to have a positive effect on the VNLGDs as well as the error performances of the codes designed by the MM-PEGA.
\end{example}

\section{Conclusions}\label{sect: conclusions}

In this paper, the metric (\ref{eqn: metric as distance}) and the metric (\ref{eqn: metric as dist and ACE pair}) were formulated firstly so that the PEG algorithm \cite{	Hu05} and the ACE constrained PEG algorithm \cite{	Vukobratovic08} were unified into one integrated algorithm, i.e., the M-PEGA.
Then, as an improvement for the M-PEGA, the MM-PEGA was proposed, which is implemented under the framework of the M-PEGA but adopts Strategy \ref{strategy: r-step M-PEGA} instead of Strategy \ref{strategy: PEG} to select the CNs.
It was illustrated that the one-edge M-PEGA is equivalent to the M-PEGA.
In addition, to calculate the multi-edge local girths used in Strategy \ref{strategy: r-step M-PEGA}, a DFS like algorithm was proposed.
%In addition, the SBM was developed to accelerate the $r$-edge M-PEGA by a factor of $m/|E|$ when the edge-trials $r > 1$.
%It's verified by the simulation results that increasing the edge-trials of the MM-PEGA is expected to have a positive effect on the cycle-structure as well as the error performance of the non-QC LDPC code designed by the MM-PEGA.
%To be specific, in Example \ref{eg: con time, ACE spectra, and BER of irreg deg15 1008 504} both the average and the maximum ACE spectra increase as the edge-trials increases and the code $B_4$ achieves the best ACE spectrum as well as the lowest BER at $E_b/N_0 = 2.8$dB, and in Example \ref{eg: reg deg3 3 codes} the VNLGD decreases as the edge-trials increases and in the best scenario the code 3M-PEGA-$3/4$ achieves a SNR-gain of $0.1$dB over the code 1M-PEGA-$3/4$ at the BER of $2\times10^{-8}$.
It was verified by the simulation results in Example \ref{eg: con time, ACE spectra, and BER of irreg deg15 1008 504} and Example \ref{eg: reg deg3 3 codes} that increasing the edge-trials of the MM-PEGA is expected to have a positive effect on the cycle-structure as well as the error performance of the non-QC LDPC code designed by the MM-PEGA.
Furthermore, to the best of our knowledge, the maximum ACE spectra reported in Table \ref{table: irreg deg15 1008 504 @ average} are the best among that spectra of the codes designed under the similar design parameters in the literature.

Meanwhile, the MM-QC-PEGA was proposed, which is implemented under the framework of the QC-PEGA \cite{Li04} but adopts a different selection strategy.
On the one hand, the MM-QC-PEGA can construct the QC-LDPC codes with any predefined valid design parameters.
On the other hand, the undetectable cycles in the QC-LDPC codes designed by the QC-PEGA \cite{Li04} and the CP-PEGA \cite{Lin08} become detectable and even avoidable in the codes designed by the MM-QC-PEGA.
It was illustrated that the $r$-edge M-PEGA is equivalent to the $r$-edge MM-QC-PEGA when the one-dimension circulant-size $N = 1$.
%However, the SBM used for accelerating the MM-PEGA cannot be applied in the MM-QC-PEGA when $N > 1$.
It was verified by the simulation results in Example \ref{eg: irreg deg15 1008 504 QC} and Example \ref{eg: reg deg3 1024 512 QC} that:
1) Compared to the QC-PEGA \cite{Li04} and the CP-PEGA \cite{Lin08}, the MM-QC-PEGA is expected to achieve better cycle-structures as well as better error performances.
2) Increasing the edge-trials of the MM-QC-PEGA is expected to have a positive effect on the cycle-structure as well as the error performance of the QC-LDPC code designed by the MM-QC-PEGA.

When designing the LDPC code with the given length, rate, and VN-degree distribution, the QC-LDPC code, which has a proper one-dimension circulant-size and is designed by the MM-QC-PEGA, may be more preferable than the non-QC-LDPC code designed by the MM-PEGA, with regard to the following three observations on the simulation results:
1) The average and the maximum ACE spectra of the QC-LDPC codes designed by the MM-QC-PEGA reported in Table \ref{table: irreg deg15 1008 504 QC} are even better than the corresponding ones of the non-QC-LDPC codes designed by the MM-PEGA in Table \ref{table: irreg deg15 1008 504 @ average}.
2) When the one-dimension circulant-size $N$ of the QC-LDPC codes designed by the MM-QC-PEGA in Table \ref{subtable: VNLGDs of 1SMCPEGQCA} and Table \ref{subtable: VNLGDs of 2SMCPEGQCA} increase, a better average and a better minimum VNLGD may be achieved.
3) Some QC-LDPC codes designed by the MM-QC-PEGA outperform the non-QC LDPC codes designed by the MM-PEGA in terms of the BER in the high SNR region, such as the best QC-LDPC code $C_{2, 2}$  in Example \ref{eg: irreg deg15 1008 504 QC} slightly outperforms the best non-QC-LDPC code $B_4$  in Example \ref{eg: con time, ACE spectra, and BER of irreg deg15 1008 504}, and the best QC-LDPC code 2M-QC-PEGA-$2^7$  in Example \ref{eg: reg deg3 1024 512 QC} slightly outperforms the best rate-$1/2$ non-QC-LDPC code 3M-PEGA-$1/2$ in Example \ref{eg: reg deg3 3 codes}.
But instead, the computational complexity of the $r$-edge M-QC-PEGA is higher than that of the $r$-edge M-PEGA.

\appendices

%\section*{Acknowledgment}
%
%
%The authors would like to thank ...

\ifCLASSOPTIONcaptionsoff
  \newpage
\fi

\bibliographystyle{ieeetr}
\bibliography{myreference}

%\end{IEEEbiography}
%
%%\begin{IEEEbiography}[{\includegraphics[width=1in,height=1.25in,clip,keepaspectratio]{fig_localGirths.pdf}}]{Michael
%%Shell}
%%.
%%.
%%\end{IEEEbiography}
%% insert where needed to balance the two columns on the last page with
%%\newpage

\end{document}